\documentclass[twocolumn,tighten,trackchanges]{aastex61}
\usepackage[us,12hr]{datetime}
\usepackage{graphicx}
\usepackage{gensymb}
\usepackage{listings}
\usepackage{color}
\usepackage{mathrsfs}
\usepackage{natbib}

\usepackage{booktabs}
\usepackage{comment}
\usepackage{morefloats}
\usepackage{sidecap}
\usepackage{grffile}
\usepackage{verbatim}
\usepackage{longtable}
\usepackage{etoolbox}
\usepackage{amsmath}
\usepackage{amssymb}
\usepackage{catchfilebetweentags}

\newcommand{\kms}{\mbox{$\rm km\,s^{-1}$}}

\newcommand{\lyalpha}{\mbox{\rm Ly$\alpha$}}

\newcommand{\hone}{\mbox{{\rm H}\,{\sc i}}}

\newcommand{\idl}{{\sc idl}}

\makeatletter
\def\CatchFBT@Fin@l#1[#2]{   \begingroup
            \makeatletter #2      \scantokens\expandafter{         \expandafter\CatchFBT@tok\expandafter{\the\CatchFBT@tok}}      \CatchFBT@IsAToken{#1}
         {\global#1\expandafter{\the\CatchFBT@tok}}
         {\xdef#1{\the\CatchFBT@tok}}      \ifx\CatchFBT@tok#1\else\global\CatchFBT@tok{}\fi
   \endgroup
}\makeatother

\graphicspath{{./figures/}}
\makeatletter
\def\input@path{{./tables/}}
\makeatother

\shorttitle{Diffuse $\lyalpha$ Nebulae at High Redshift}
\shortauthors{Xue et al.}

\begin{document}
\title{The Diversity of Diffuse $\lyalpha$ Nebulae around Star-Forming Galaxies at High Redshift}

\author[0000-0001-7689-9305]{Rui Xue}
\affiliation{Department of Physics and Astronomy, Purdue University, 525 Northwestern Avenue, West Lafayette, IN 47907, USA}

\author{Kyoung-Soo Lee}
\altaffiliation{Visiting Astronomer, Kitt Peak National Observatory, National Optical Astronomy Observatory, which is operated by the Association of Universities for Research in Astronomy (AURA) under cooperative agreement with the National Science Foundation.}
\affiliation{Department of Physics and Astronomy, Purdue University, 525 Northwestern Avenue, West Lafayette, IN 47907, USA}

\author{Arjun Dey}  
\affiliation{National Optical Astronomy Observatory, 950 N. Cherry Avenue, Tucson, AZ 85719, USA}

\author{Naveen Reddy}
\altaffiliation{Sloan Research Fellow}
\affiliation{Department of Physics and Astronomy, University of California, Riverside, 900 University Avenue, Riverside, CA 92521, USA}

\author{Sungryong Hong}
\affiliation{Department of Astronomy, University of Texas at Austin, 2515 Speedway, Stop C1400, Austin, TX 78712, USA}

\author{Moire K.~M. Prescott}
\affiliation{Department of Astronomy, New Mexico State University, P.O. Box 30001, Las Cruces, NM 88001, USA}

\author{Hanae Inami}
\affiliation{National Optical Astronomy Observatory, 950 N. Cherry Avenue, Tucson, AZ 85719, USA}

\author{Buell T. Jannuzi}
\affiliation{Steward Observatory, University of Arizona, 933 N Cherry Ave, Tucson, AZ 85721, USA}

\author{Anthony H. Gonzalez}
\affiliation{Department of Astronomy, University of Florida, 211 Bryant Space Science Center, Gainesville, FL 32611, USA}

\begin{abstract}
We report the detection of diffuse Ly$\alpha$ emission, or Ly$\alpha$ halos (LAHs), around star-forming galaxies at $z\approx3.78$ and $2.66$ in the NOAO Deep Wide-Field Survey Bo\"otes field.
Our samples consist of a total of $\sim$1400 galaxies, within two separate regions containing spectroscopically confirmed galaxy overdensities. They provide a unique opportunity to investigate how the LAH characteristics vary with host galaxy large-scale environment and physical properties.
We stack Ly$\alpha$ images of different samples defined by these properties and measure their median LAH sizes by decomposing the stacked Ly$\alpha$ radial profile into a compact galaxy-like and an extended halo-like component.
We find that the exponential scale-length of LAHs depends on UV continuum and Ly$\alpha$ luminosities, but not on Ly$\alpha$ equivalent widths or galaxy overdensity parameters.
The full samples, which are dominated by low UV-continuum luminosity Ly$\alpha$ emitters ($M_{\rm UV} \gtrsim -21$), exhibit LAH sizes of 5$\,-\,6\,$kpc. 
However, the most UV- or Ly$\alpha$-luminous galaxies have more extended halos with scale-lengths of 7$\,-\,9\,$kpc.
The stacked Ly$\alpha$ radial profiles decline more steeply than recent theoretical predictions that include the contributions from gravitational cooling of infalling gas and from low-level star formation in satellites. 
On the other hand, the LAH extent matches what one would expect for photons produced in the galaxy and then resonantly scattered by gas in an outflowing envelope.
The observed trends of LAH sizes with host galaxy properties suggest that the physical conditions of the circumgalactic medium (covering fraction, \hone\ column density, and outflow velocity) change with halo mass and/or star-formation rates.
\end{abstract}

\section{Introduction}\label{intro}

The circumgalactic medium (CGM) encodes the details of two main physical processes that shape how galaxies form and evolve: namely, the gas accretion that fuels star formation and the resulting feedback.
Constraining the spatial distribution and dynamical state of the CGM around high-redshift galaxies is thus of critical importance to test our current theoretical framework of galaxy formation.

However, the diffuse nature of the CGM poses a major observational challenge. 
Direct 21\,cm imaging of atomic gas in the CGM is beyond the capability of current instruments \citep{Carilli:2004es}.
While absorption sightlines can probe the CGM \citep[e.g.,][]{Steidel:2010go, Tumlinson:2013cl}, inferring the gas distribution from them is nontrivial due to the discrete sampling of bright background sources.
Diffuse Ly$\alpha$ emission, or Ly$\alpha$ halos (LAHs), around high-redshift galaxies, may open a new avenue for the CGM study:
Ly$\alpha$ photons -- which are presumably produced copiously at sites of star formation -- can be resonantly scattered by neutral hydrogen gas out to large galactocentric distances; as a result, the \lyalpha\ emission is expected to appear more extended than the rest-frame UV continuum. 
With sophisticated and self-consistent models of Ly$\alpha$ radiative transfer modeling, it is possible to follow how Ly$\alpha$ photons propagate through simulated interstellar, circumgalactic, and intergalactic media.
Comparisons between models and observations may provide invaluable insight into the physical properties of the CGM neutral gas \citep{Zheng:2011kz,Dijkstra:2012ju,Verhamme:2012kb,Lake:2015gm}.

At high redshift, there is growing evidence for the presence of LAHs around high-redshift star-forming galaxies \citep[][]{Hayashino:2004ke,Rauch:2008jy, Steidel:2011jk, Matsuda:2012fp, Momose:2014fe, Momose:2016cu}, although a few studies find only marginal detections in their samples \citep[e.g.,][]{Feldmeier:2013fx, Jiang:2013cm}.
Recent \lyalpha\ and UV observations also suggest that \lyalpha-emitting galaxies in the local universe ubiquitously produce large-scale halos of scattered \lyalpha\ emission \citep{Hayes:2015kv}.
In spite of the observational progress, it is still unclear what the dominant power source is that produces extended Ly$\alpha$ emission around galaxies. 
While star formation inside the host galaxy is likely to produce Ly$\alpha$ photons that scatter through the medium, spatially extended Ly$\alpha$ emission may also originate from widespread low-level star-formation activity or from the cooling radiation of collisionally heated infalling gas, beyond the regions revealed by rest-frame UV or IR emission.
The uncertainty of ionizing source distributions from these auxiliary mechanisms remains a major obstacle to establishing a direct relation between the appearance of LAHs and the physical properties of the CGM.

On the other hand, the variations of LAH characteristics measured in different high-redshift galaxy samples clearly require physical explanations.
Based on different sample selections, some studies find more extended halos \citep[hereafter S11]{Steidel:2011jk} than others \citep{Feldmeier:2013fx, Momose:2014fe, Wisotzki:2016hw}.
The difference could be elaborated by considering various scenarios:
the spatial and velocity structures of the interstellar medium (ISM) and CGM likely change as a function of galaxy properties and host halo \citep[e.g.,][]{Jones:2013ed,Williams:2015gi}, which can alter $\lyalpha$ photon propagation and the LAH surface brightness profile;
the \lyalpha\ spatial extent may also be strongly related to the surrounding megaparsec-scale environments \citep[]{Matsuda:2012fp}, provided that the CGM gas structure or the ionizing source distribution outside the host galaxy (i.e., satellite galaxies and gas accretion) is associated with the dark matter distribution \citep[e.g.,][]{Laursen:2007kl,Zheng:2011kz}.
Identification of a clear trend of how LAH characteristics depend on galaxy properties may provide a promising avenue to elucidate the origin of the LAH phenomenon and its connection with the CGM.

In this paper, we present new LAH measurements based on two of the largest spectroscopic and photometric samples of high-redshift star-forming galaxies. 
We focus on examining how LAH characteristics change with galaxy UV continuum and \lyalpha\ luminosities, and large-scale environment.
This is achieved in practice by stacking Ly$\alpha$ and UV continuum images of subsets of galaxies binned by a specific property.
The paper is organized as follows. 
In Section\,\ref{sec:datasample}, we describe the data sets, sample selection, and other notable facts about our survey fields. 
In Section\,\ref{sec:method}, we present the methodology adopted for the image stacking analysis and measurements of the LAH properties. 
In Section\,\ref{sec:measure_overdensity}, we discuss the characterization of galaxies' large-scale environments, one of the key parameters we explore in the LAH size dependence. 
Our main findings are described in Section\,\ref{sec:results}.
In Section\,\ref{sec:discussion}, we discuss the implications of our results in the context of  recent theoretical predictions. 
Finally, a summary of our results and conclusions is given in Section\,\ref{sec:summary}. 

Throughout this paper, we use the WMAP7 cosmology $(\Omega, \Omega_\Lambda, \sigma_8, h) = (0.27, 0.73, 0.8, 0.7)$ from \citet{Komatsu:2011in}. Distance scales are presented in units of comoving megaparsecs unless noted otherwise. 
All magnitudes are given in the AB system \citep{Oke:1983jp}. 

\section{Data and Samples of Galaxies in Diverse Environments}\label{sec:datasample}

We consider two galaxy samples for our investigation. These samples consist of star-forming galaxies at $z\approx3.78$ and $2.66$, and populate two non-overlapping regions in the Bo\"otes field of the NOAO Deep Wide-Field Survey \citep[NDWFS;][]{Jannuzi:1999wu}.
Here, we briefly describe the samples and their characteristics. 

\begin{figure*}[!htbp]
\begin{minipage}{0.35\linewidth}
\includegraphics[width=1.0\linewidth]{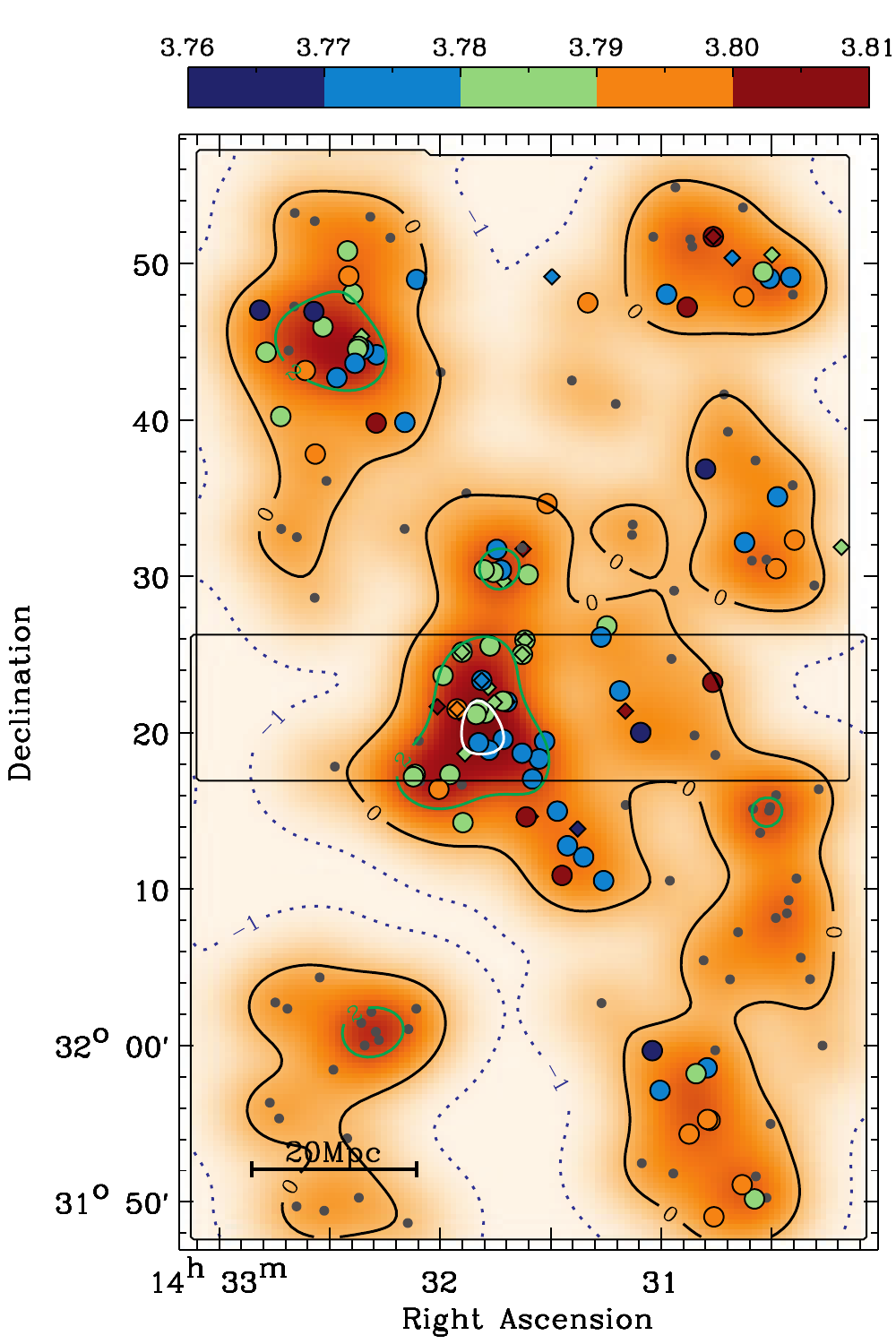}
\end{minipage}
\begin{minipage}{0.64\linewidth}
\includegraphics[width=1.0\linewidth]{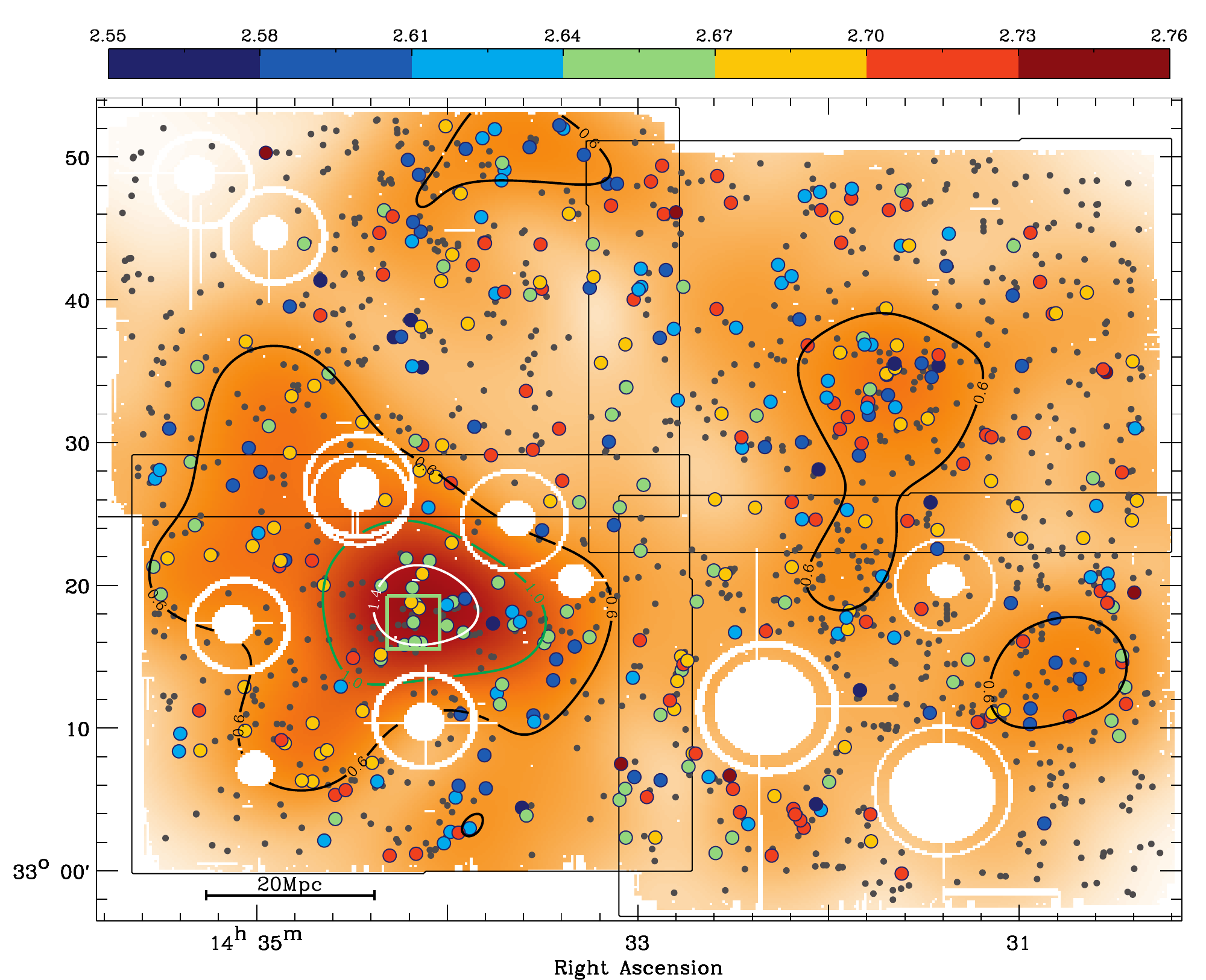}
\end{minipage}
\caption{\label{fig:pcflab_env}
Left: the surface overdensity map of the LAEs detected at $z\approx3.78$ in the PCF field, shown in color scale and contours.
The positions of photometric LAEs are shown as gray dots, while spectroscopic sources are color coded in redshift (filled circles and diamonds for LAEs and LBGs, respectively).
The coverages of the PCF-N and -S pointings are indicated by solid black boxes.  
As described in Section\,\ref{sec:laeoverd}, the overdensity map is derived from the LAE surface density, which is constructed by smoothing the LAE positions with a 2D Gaussian kernel of ${\rm FWHM}=4\farcm8$, or 10\,Mpc at $z=3.78$.
Right: the second moment map of the 3D overdensity of the spectroscopic LAEs in the LAB field at $z\approx2.66$, shown in color scale.
Gray dots represent the photometric LAEs, and the spectroscopically confirmed LAEs at $z=2.569-2.737$ are indicated by filled circles, color coded with redshift.
The boundaries of four LAB pointings are marked with solid black boxes, and the masked regions (due to image saturation and other artifacts) are in white.
The 3D overdensity of the LAB field is estimated by smoothing the distribution of the spectroscopic LAEs in the comoving volume with a 3D Gaussian kernel of ${\rm FWHM}=20$\,Mpc.
In both panels, the comoving distance scale at corresponding sample redshift is indicated at their bottom left corners.
}
\end{figure*}

\subsection{Galaxies around PC~217.96+32.3 at $z\approx3.78$}\label{pcf_data}

The first sample consists of galaxies at $z\approx3.78$, and will be referred to as the ``PCF'' (protocluster field) sample hereafter. 
These galaxies are distributed over a $1\fdg2 \times 0\fdg6$ contiguous region located at the southern edge of the Bo\"otes field (center: $\alpha=217\fdg86$, $\delta=32\fdg33$).
The optical data are taken with the Mosaic 1.1 wide-field imaging camera on the Mayall 4m telescope of the Kitt Peak National Observatory. 
The field consists of two adjacent pointings with a $9\arcmin$ overlap in the north--south direction (see Figure\,\ref{fig:pcflab_env}).
The northern pointing (PCF-N) coincides with an NDWFS subfield named NDWFSJ1431+3236, while the southern pointing (PCF-S) lies outside the NDWFS boundary. 
The PCF-N imaging includes the NDWFS data\footnote{\url{http://www.noao.edu/noao/noaodeep/DR3/dr3-data.html}} complemented by our new observations taken in 2012 May. 
The PCF-S observations are taken in 2014 May and June, closely matching the sensitivities of the PCF-N data. 
The entire PCF field is observed with four NOAO filters: three broadband filters ($B_W$, $R$, $I$; $\lambda_{\rm{cen}}=4222, 6652, 8118$\,\AA) and one narrowband filter {\it WRC4}. 
The {\it WRC4} filter is designed to sample C\,{\sc iv} emission in Wolf--Rayet stars (KPNO filter no.\,k1024).
In the KPNO 4m $f$/3.1 prime-focus corrector beam, it has a central wavelength of 5819\,\AA\ and a FWHM of 42\,\AA, which can sample Ly$\alpha$ emission at $3.775<z<3.810$ (i.e., 27 comoving Mpc along the line of sight). 
At the same redshift, the three broadband central wavelengths correspond to rest-frame wavelengths of 882, 1390, and 1697\,\AA, respectively. 
The details of these observations are presented in \citet{Lee:2014gv} and \citet{Dey:2016dl}. 

The galaxies in the PCF sample are selected as \lyalpha\ emitters (LAEs) or Lyman break galaxies (LBGs). 
The LAE selection requires blue narrowband-to-broadband colors (${\it WRC4}-R$), which is designed to isolate galaxies that have an excess \lyalpha\ emission in the {\it WRC4} band.
The LBG candidates are selected by applying a Lyman break color selection technique to the $B_W R I$ data \citep[e.g.,][]{Steidel:1999ee,Giavalisco:2004et,Bouwens:2007jj,Lee:2011dw}. The technique is designed to identify UV-bright star-forming galaxies at $3.3<z<4.3$, that show strong Lyman break at rest-frame $\lambda \leq 1216$\,\AA\ (between the $B_W$ and $R$ bands).
The adopted LAE color criteria are
\begin{eqnarray}
\label{eq:cc-lae}
({\it WRC4}-R) <-0.8 & ~~\cap~~ & {\rm S/N}({\it WRC4})\geq 7 \nonumber \\
\cap~[ (B_W-R)>1.8 & \cup & {\rm S/N}(B_W) < 2 ],
\end{eqnarray}
and the LBG color criteria are
\begin{eqnarray}
\label{eq:cc-lbg}
(B_W-R)> 3~(R-I)+1.74~\cap~(B_W-R)\geq 1.8 \nonumber \\
 (R-I) \geq -0.54~\cap~{\rm S/N}(R) \geq 3~\cap~S/N(I)\geq 7.
\end{eqnarray}

Follow-up spectroscopy was carried out, targeting a subset of the photometric LBG and LAE samples, the results of which are described in \citet{Lee:2013gk} and \citet{Dey:2016dl}. 
Briefly, of the 165 LAE candidates, 100 are observed, only two of which are identified as [O\,{\sc ii}] emitters at low redshift. Counting null detections and two interlopers as failures, the spectroscopic success rate for LAEs is found to be 89\%, suggesting that the majority of unobserved LAEs also lie at the expected redshift of $z\approx3.78$.
None of the confirmed LAEs shows evidence of strong active galactic nucleus (AGN) activity in their spectra (i.e., broad/strong lines of high ionization species).
Hence, we include the entire photometric sample of LAEs (excluding the two confirmed interlopers) for our analyses. On the other hand, the LBG candidates span a much wider redshift range \citep[$z=3.7\pm0.4$,][]{Lee:2013gk}, and therefore we only include LBGs that are spectroscopically confirmed to lie at the same range as LAEs: $z_{\rm{spec}}=3.775$\,--\,$3.810$. 
Table~\ref{tab:pworks} includes the numbers of LAEs and LBGs in the PCF field used for the present analyses.
 
Within the PCF field lie two massive protoclusters at $z=3.78$, which we collectively dub PC~217.96+32.3. One is located near the field center at $(\alpha,\,\delta)=(217\fdg91,\,+32\fdg35)$ and contains 39 spectroscopically confirmed members at $z=3.774$\,--\,$3.790$, while the other is located at the northeastern corner of the field, $(\alpha,\,\delta)=(218\fdg15,\,+32\fdg75)$, with 16 confirmed members at $z=3.775$\,--\,$3.796$.
\citet{Dey:2016dl} estimated the total masses enclosed to be $\approx 10^{15}M_\odot$ and $\approx 6\times 10^{14}M_\odot$ for the central and northeastern structures, respectively.
Both structures represent extremely rare high-overdensity regions, and are expected to evolve into massive galaxy clusters by the present-day epoch.
Further discussion on the environment measurements in and around these structures is given in Section\,\ref{sec:measure_overdensity}. 

\subsection{Galaxies around LABd05 at $z\approx2.66$}\label{lab_data}

Our second sample consists of LAEs around a giant Ly$\alpha$ nebula or Ly$\alpha$ blob (LAB) at $z=2.656$  within the Bo\"otes field at $(\alpha,\,\delta)=(218\fdg546,\,33\fdg291)$. The nebula, known as LABd05, has a line luminosity $L_{\rm{Ly}\alpha}\approx 1.7\times 10^{44}~\rm{erg~s}^{-1}$, and is spatially extended to a radius of at least 15\arcsec, or a projected physical distance of $\approx120$~kpc \citep{Dey:2005dl}.
The optical data set includes the broadband NDWFS $B_WRI$ data and the {\it IA445} intermediate-band filter data ($\lambda_{\rm{cen}}=4458\rm{\AA}, \Delta \lambda =201\rm{\AA}$) taken with the SuprimeCam imager on the Subaru telescope. The {\it IA445} filter can sample the redshifted Ly$\alpha$ emission at $z=2.569$\,--\,$2.737$, corresponding to the line-of-sight comoving distance of $\approx190$~Mpc. Four adjacent pointings are taken in each filter, covering a 1\,deg$^2$ contiguous region (see Figure\,\ref{fig:pcflab_env}). 

The adopted LAE selection criteria are
\begin{eqnarray}
\label{eq:cc-lab-lae}
({\it IA445}-B_W) &<&-0.5~\cap~{\rm S/N}({\it IA445})>7\nonumber \\
~\cap~B_W-R &\leq& 0.8.
\end{eqnarray}
The first criterion isolates galaxies with a strong emission line falling into the {\it IA445} band, while the last one is intended to eliminate contaminant populations, namely [O\,{\sc ii}] emitters at $z\approx0.2$ or higher-redshift star-forming galaxies at $z\gtrsim3$. More details of the Subaru observation, data reduction, and candidate selection are provided in \citet{Prescott:2008jg}.
The selection yields 1336 LAE photometric candidates in regions free of image artifacts (after the exclusion of known contaminants identified in the follow-up spectroscopy, see below).
The LAE candidates in this field, which we refer to as the LAB sample, are expected to have the rest-frame equivalent widths of $\rm{EW}_0\gtrsim 50$\,\AA, and thus are typically stronger line emitters than the LAEs in the PCF field ($\geq 20$\,\AA).

Follow-up spectroscopy was performed on a subset of LAEs with the Hectospec instrument, a bench-mounted fiber spectrograph on the 6.5m MMT.
The details of target selections and spectroscopic observations are presented in \citet{Hong:2014im}. 
The Hectospec observation consisted of seven pointings, whose center positions are offset up to 30\arcmin\ from one another in the RA direction to cover the entire field. 
Therefore, the fraction of photometric LAEs observed by Hectospec is not completely uniform across the field due to the technical limitations of assigning fibers to sources far from pointing center. As a result, the middle two-thirds of the LAB field is more densely populated by spectroscopic LAEs than the eastern and western ends.

The observations confirmed 429 LAEs at $z=2.569$\,--\,$2.737$, corresponding to $\approx87$\% of the LAE candidates with spectroscopic redshifts, suggesting a relatively clean photometric sample.
The imaging field of the LAB sample also contains at least one overdensity region in the immediate vicinity of LABd05. Multiple companions are discovered around LABd05, including an AGN and an LBG within the nebula \citep{Dey:2005dl}. 
Deep $HST$ imaging further revealed a population of compact low-luminosity galaxies in its vicinity, and \citet{Prescott:2012kp} argued this as evidence that LABd05 is a site of an ongoing group formation. Their speculation is corroborated by the MMT spectroscopy, which identified 168 LAEs within $\Delta z=\pm0.02$ of the LABd05. 
In Section\,\ref{sec:measure_overdensity}, we further quantify the LAE-defined local environments within the LAB field. 

\floattable
\begin{deluxetable}{cccccc}[htb!]
\tabletypesize{\footnotesize}
\tablecolumns{8} 
\tablecaption{Sample Summary and Comparisons with the Literature\label{tab:pworks}}
\tablewidth{0pt} 
\tablehead{
\multicolumn{1}{c}{Field}&
\multicolumn{1}{c}{Selection}&
\multicolumn{1}{c}{Sample Size\tablenotemark{a}}&
\multicolumn{1}{c}{Redshift\tablenotemark{b}}&
\multicolumn{1}{c}{$\delta$ kernel\tablenotemark{c}}&
\multicolumn{1}{c}{Reference}
}
\startdata
PCF & LAE& 163 & $3.775-3.810$ & Gaussian & This Work \\
- & LBG & 21 & $3.775-3.810$ & $\sigma=2\farcm0$(4.3\,Mpc) & \nodata\\
LAB & LAE & 1336 & $2.569-2.737$ & $\sigma=4\farcm8$(8.6\,Mpc) & \nodata \\
\midrule
HS1549+195 & LBG & 27 & $2.802-2.875$ & \nodata & \citet{Steidel:2011jk} \\
HS1700+643 & LBG & 43 & $2.266-2.340$ & \nodata & \nodata \\
SSA22a & LBG & 22 & $3.063-3.129$ & \nodata & \nodata \\
\midrule
GOODS-N+SDF+&LAE &  2128 & $3.062-3.126$ & Gaussian  & \citet{Matsuda:2012fp} \\
SXDS+extended SSA22a  & LBG &  24 & $3.062-3.126$ & $\sigma=1\farcm5$(2.9\,Mpc) & \\
\midrule
SXDS & LAE & 316 & $3.106-3.167$ & \nodata & \citet{Momose:2014fe}\\
\nodata & LAE & 100 & $3.663-3.720$ & \nodata & \nodata \\
\nodata & LAE & 397 & $5.655-5.753$ & \nodata & \nodata \\
\nodata & LAE & 119 & $6.510-6.619$ & \nodata & \nodata \\
COSMOS+GOODS & LAE & 3556 & $2.145-2.222$ & Tophat & \citet{Momose:2016cu} \\
+SSA22+SXDS & & & & $r=10\arcmin$(16.2\,Mpc) & \nodata \\
\enddata
\tablenotetext{a}{The sizes of our samples here represent the number of galaxies selected for image stacking.
For the PCF and LAB samples, 98 and 429 galaxies have been confirmed spectroscopically within the expected redshift ranges, respectively.
}
\tablenotetext{b}{The photometric redshift range is derived by filter half-power points.}
\tablenotetext{c}{The smoothing kernel used to derive overdensity (FWHM in parentheses).}
\end{deluxetable}

\floattable
\begin{deluxetable}{ccccccc}[htb!]
\tabletypesize{\scriptsize}
\tablecolumns{5} 
\tablecaption{Imaging Data and PSF Summary \label{tab:psfprop}}
\tablewidth{0pt}
\tablehead{
\colhead{Data set/} & 
\colhead{Pointing Name\tablenotemark{a}} & 
\colhead{{\it WRC4}}&
\colhead{{\it IA445}}&
\colhead{$B_{W}$}&
\colhead{$R$}&
\colhead{$I$}\\
\colhead{Redshift}&
\colhead{} &
\colhead{5820\,\AA/42\,\AA\tablenotemark{b}} &
\colhead{4458\,\AA/201\,\AA} &
\colhead{4222\,\AA/1275\,\AA} &
\colhead{6652\,\AA/1511\,\AA} &
\colhead{8118\,\AA/1915\,\AA} 
}
\startdata
PCF/ & PCF-N (J1431p3236) & 0\farcs88 &\nodata & (0\farcs97)  & 0\farcs95 & 0\farcs77 \\
$z\approx3.78$ & PCF-S & 1\farcs00 & \nodata & (1\farcs28)  & 0\farcs97 & \underline{1\farcs10} \\
\midrule
& NDWFS1 (J1434p3311)& \nodata & 0\farcs73 & 1\farcs00  & 1\farcs17 & (0\farcs95) \\
LAB/ & NDWFS4 (J1431p3311)&\nodata & 0\farcs80& 1\farcs13  & 1\farcs05 & (1\farcs10) \\
$z\approx2.66$ & NDWFS5 (J1434p3346)& \nodata & 0\farcs95 & 0\farcs97  & 0\farcs85 & (1\farcs14) \\
& NDWFS6 (J1431p3346)& \nodata & 1\farcs00 & \underline{1\farcs37}  & 1\farcs14 & (1\farcs00)\\
\enddata
\tablecomments{
The PSF sizes (FWHMs) are measured from the {\sc PSFEx} models (see descriptions in Section\,\ref{sec:psf}).
We use parentheses to denote the archival NDWFS data that were not included in our image analyses.
The underlined values correspond to the worst seeing data in individuallly used data sets.
}
\tablenotetext{a}{The corresponding NDWFS subfield names are given in parentheses, with the prefix ``NDWFS" omitted.}
\tablenotetext{b}{The filter central wavelength/FWHM, $\lambda_{\rm{cen}}/\Delta \lambda$.}
\end{deluxetable}

\section{Testing for the Presence of Diffuse Emission around Galaxies}\label{sec:method}

Our primary goal is to test for the presence of LAHs around galaxies in our samples.
Because the \lyalpha\ sensitivity of our observations is lower than the expected LAH surface brightness, we determine average light profiles by stacking the \lyalpha\ images of many galaxies. In practice, this is a challenging task because diffuse emission can be easily mimicked by observational factors, such as image misregistration,  point-spread functions (PSF), or imperfect sky subtraction that varies from field to field. Some of these effects are discussed extensively in \citet{Feldmeier:2013fx}. 
In this section, we describe the procedures for our image analyses and various tests we performed to quantify uncertainties. 

\subsection{Image Registration, PSF, and PSF Matching}\label{sec:psf}

Two critical elements for robust stacking analyses are image registration and PSF homogenization. 
Both misregistration and varying PSFs can artificially broaden a stacked galaxy light profile.
We eliminate these possibilities by employing the procedures described below.

For each exposure of a given pointing, we update the astrometry (employing the {\tt iraf}\footnote{IRAF is distributed by the National Optical Astronomy Observatory, which is operated by the Association of Universities for Research in Astronomy (AURA) under cooperative agreement with the National Science Foundation.} task {\tt mscred msccmatch}), using stars identified in the Sloan Digital Sky Survey DR7 Catalog, and reproject it to a common tangent point using a sinc interpolator with a pixel scale of $0\farcs258$~pixel$^{-1}$. 
The same procedure is repeated for all frames from new observations or the NDWFS archive.
The typical RMS in the astrometric solution is 0\farcs05\,--\,0\farcs08.
For each pointing per band, a final mosaicked image is created as a weighted average of all frames, with weights inversely proportional to the variance of sky noise measured in the reprojected frames.

Next, we determine the PSFs of individual images using two different approaches. 
In the first approach, we adopt a procedure similar to that outlined by \citet{Feldmeier:2013fx}. We first run {\sc SExtractor} \citep{Bertin:1996ww} to create the source catalog of each image. A list of relatively bright but unsaturated stars is created based on the compactness parameter ({\tt CLASS\_STAR$>$0.95}), and then cross-referenced with the Guide Star Catalogs v2.3.2 \citep[GSC, {\tt CLASS}=0:][]{Lasker:2008ju}. Based on the catalogs and visual inspection, sources with blending issues or with a companion within 5$\arcsec$ are removed from the list.  
We measure the radial profile of each star by azimuthally averaging in bins of annuli, and normalize the profile at 2\arcsec\ from center. 
Then we take a median average of all normalized star radial profiles to derive the PSF out to 4\arcsec. 
Independently of this procedure, we also measure the large-scale ``wings'' of the PSF from the GSC stars that are saturated in our data. We first mask saturated pixels and determine the centroids by fitting a two-dimensional Gaussian function to the unmasked pixels. The outer radial profile is then measured at 2\arcsec--\,6\arcsec, and is combined with the inner profile by matching the amplitude at 3\arcsec.

In an alternative approach, we use a software {\sc PSFEx} \citep{Bertin:2011wm} to derive PSF models. {\sc PSFEx} uses the information supplied by a {\sc SExtractor} source catalog to iteratively reject sources that are saturated, contaminated, or extended, and then automatically selects point source candidates to construct a non-parametric PSF model. Specifically, for each image, we supply an input catalog including the sources with  {\tt ELLIPICITY$<$0.1} and {\tt SNR\_WIN$>$30}, where {\tt ELLIPICITY} and {\tt SNR\_WIN} are SExtractor parameters defined as source ellipticity and window-based signal-to-noise ratio (S/N), respectively.
{\sc PSFEx} will reject any saturated sources and does not use the broad light profile of the saturated object to derive PSF as our first method, however, a larger number of point sources are usually included due to the efficient automated selection criteria.
In the {\sc PSFEx}-based method, we can model the PSF spatial variation within each image using second-order polynomial components.
But the zeroth-order component is found to adequately describe all PSFs in our stacking analyses (see details below) and therefore is adopted in this study.

\begin{figure*}
\centering
\includegraphics[width=0.80\paperwidth]{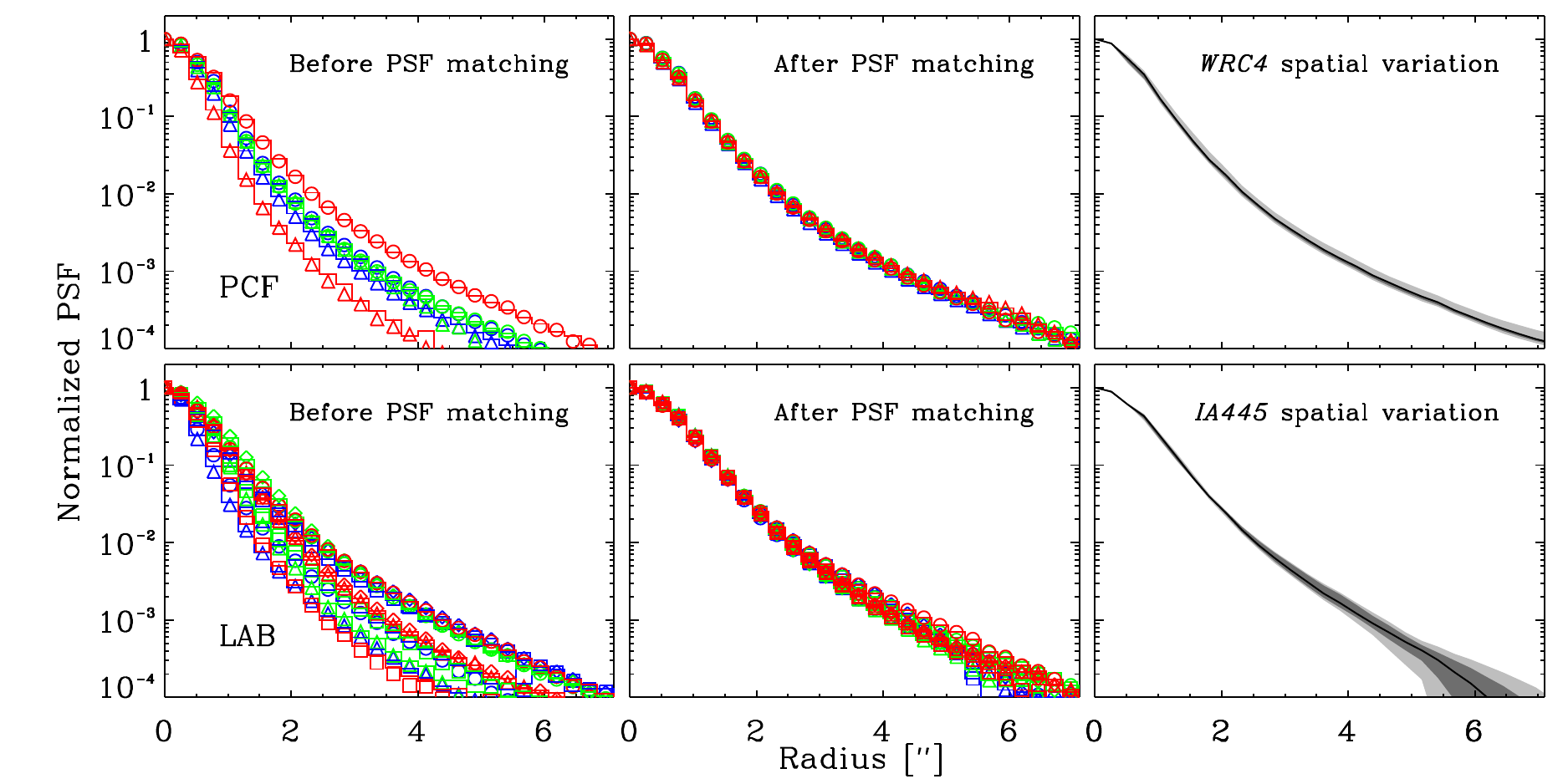}
\caption{\label{fig:radprofile_all_psf}
Radially averaged light profiles of Galactic stars are measured to determine the point-spread functions of the PCF (top) and LAB (bottom) image tiles.
The left panels show the PSF measurements of original images.
The colors of open histograms and symbols represent different pointings and passbands.
As described in Section\,\ref{sec:psf}, two different methods are employed to determine the PSFs, which are shown as histograms (GSC-based) and symbols (PSFEx-based) in each panel.
Both methods return very similar results, and allow us to robustly characterize the PSF effects on observed light profiles of galaxies within a dynamic range of  up to three orders of magnitude.
In the middle panels, we show the estimated PSFs of all image tiles after homogenizations, suggesting that they are in good agreement out to $\approx 6$\arcsec.
In the right panels, we present the 10\%\,--\,90\% and 25\%\,--75\% ranges of position-dependent {\sc PSFEx} models at sample galaxy locations as the light and dark gray shades, with the black lines showing the median profiles.
They are derived from the PSF-homogenized NB images.
}
\end{figure*}

In the left panels of Figure~\ref{fig:radprofile_all_psf}, we compare the stellar PSFs determined by the GSC-based  (histograms) and PSFEx-based (open symbols) methods, with colors representing different pointings and passbands.
The results illustrate large PSF differences within our data sets, but two methods are evidently in excellent agreement  for a given image. 
For simplicity, we adopt the PSFEx-based PSF models for all analyses in this study. 
The PSF FWHM of each image is listed in Table~\ref{tab:psfprop}.

We homogenize the PSFs by convolving all images in each field to the reference PSF, which is taken as that of the worst seeing data in the data set. 
For example, the reference PSF for the PCF sample is from the $I$-band image of the PCF-S pointing (see Table~\ref{tab:psfprop}).
A convolution kernel is derived by comparing the circularized PSF\footnote{A circularized PSF is generated from the radially medium-averaged PSF image produced by {\sc PSFEx}, leading to a higher S/N at the outer profile at the expense of simplifying the PSF into a 1D function. Using circularized PSFs will produce a noiseless convolution kernel.} of each image with the reference using an \idl\ deconvolution routine {\sc max\_likelihood}.  The kernel is then used to convolve the image using the \idl\ {\sc convol} routine.
To verify the results, we repeat the {\sc PSFEx}-based procedure on the PSF-homogenized images using the same set of stars. 
The derived PSF models are presented in the middle panels of Figure\,\ref{fig:radprofile_all_psf}, showing that the PSFs within a given data set are now in reasonable agreement with one another out to $6\arcsec$. 
The FWHMs of homogenized PSFs are 1\farcs10 and 1\farcs37 for the PCF and LAB sample, respectively. 
In the right panels of Figure\,\ref{fig:radprofile_all_psf}, we present the estimated PSF spatial variation in the PSF-homogenized NB images by including the first- and second-order {\sc PSFEx} model components.
Because the variations are mild as illustrated,  adopting spatially varying PSFs for our subsequent analyses will have little impact on our main conclusions.

\subsection{Measuring Light Profiles of Galaxies}
\label{sec:defaultstack}

Having homogenized all of the images to a common PSF within $\approx6\arcsec$, we adopt the following procedure for image stacking preparation and galaxy light profile measurements.

First, we make a $1\arcmin\times1\arcmin$ cutout image centered on each galaxy.  {\sc SExtractor} is run on the image twice to detect all objects down to 1.5$\sigma$ and 2$\sigma$ with {\sc detect\_minarea}$=5$pixel and {\sc back\_size}$=30\arcsec$. 
We apply a Gaussian filter with an FWHM of 2 pixels for improving  detection of extended faint objects. 
Using the {\sc SExtractor} segmentation maps, we mask all 2$\sigma$ detection pixels belonging to interlopers (i.e., sources not at center) and expand individual mask regions to adjacent 1.5$\sigma$ regions for a more robust exclusion of low-level contaminants.
Cutout images are then oversampled to 0\farcs086/pixel (i.e., one-third of the original pixel size) using a nearest-neighbor interpolation, with the object centroid located at the center pixel of the resampled image. 
Finer resampling ensures that image stacking can be conducted without introducing further broadening due to sub-pixel centroid uncertainties. Finally, we ``repair'' the masked regions as follows. 
For each pixel within mask regions, we determine its distance from the galaxy center. 
Then, we replace its value with the median value of the unmasked pixels at a similar projected galactocentric distance and further add pseudo-noise commensurate with the measured cutout background noise. 
This additional step ensures that all objects contribute equally to each pixel of the stacked image.
Generally, fewer than 20\% of the pixels in the cutout images are masked. Whenever more than 20\% of the area is masked out, we do not include the galaxy in the stacking procedure.

Following the method detailed in Appendix\,\ref{sec:flya}, we create Ly$\alpha$ and line-free continuum images at rest-frame $\sim$1220\,\AA\ using two bands that sample both Ly$\alpha$ and UV continuum emission. We refer to them as the narrow- and broadbands\footnote{
The NB and BB here are {\it WRC4} and $R$ bands for the PCF sample, and {\it IA445} and $B_W$ bands for the LAB sample, respectively.
Although {\it IA445} is an intermediate-band filter, we refer to it as NB for notational convenience.} (NB and BB).
The median $3\sigma$ surface brightness sensitivities of individual resampled cutout Ly$\alpha$ images are $6.3\times10^{-18}$ and $13.2\times10^{-18}$\,erg\,s$^{-1}$\,cm$^{-2}$\,arcsec$^{-2}$ in the PCF and LAB sample, respectively. 
For the galaxies at $z\approx3.78$ or $z\approx2.66$, we use the $I$- or $R$-band images to estimate their UV continuum at rest-frame $\sim$1700\,\AA\ (referred to as the continuum band, or CB).
The median $3\sigma$ surface brightness sensitivities of cutout images reach $9.5\times10^{-31}$ and $7.2\times10^{-31}$\,erg\,s$^{-1}$\,cm$^{-2}$\,arcsec$^{-2}$\,Hz$^{-1}$, respectively.

To create a stacked image for a given galaxy sample, we exclude the galaxies that are within $5\arcsec$ from a bright source, within $1$\arcmin\ from the edges of the science image, or of the masked regions near bright saturated stars (shown in Figure\,\ref{fig:pcflab_env}). Then we simply take a pixel-to-pixel median of the resampled, re-centered cutout images. 
This procedure excludes $\approx$13\% and $\approx$2\% of the LAB sample and PCF sample, respectively. The higher level of source removal in the LAB sample is a result of the fact that the Subaru {\it IA445} band images contain a larger number of artifacts (CCD bleeding and diffraction spikes around bright stars; see Figure \ref{fig:pcflab_env}), and thus require more aggressive masking in order to avoid contamination.

The cutout images likely still contain emission from projected nearby sources below our adopted detection threshold. They may introduce negative or positive bias regions in \lyalpha\ cutout images due to imperfect continuum subtraction. However, assuming that these undetected objects are randomly distributed in individual cutout images without the presence of clustering, their contribution should be similar across all pixels of the stacked image. 
To create a ``contamination-free'' stacked image, we estimate the effective background using an annular region from 6\arcsec\ to 10\arcsec\ and subtract its median value from the stacked image.  
This procedure removes the bias introduced by undetected objects, but limits our ability to identify any diffuse emission beyond the central region 12\arcsec\ in diameter (corresponding to 88 and 98\,kpc in the angular distance, for our samples at $z=3.78$ and $z=2.66$, respectively). 
The choice of the annular region size will be further discussed in Appendix\,\ref{sec:stack_effect}.

The radial surface brightness profiles are measured by azimuthally averaging the pixel values in successive bins of annuli from source center. We find that the median and mean averaging within annuli give consistent results in all cases of our study. 
Here, we adopt median averaging results as observed light profiles.  
To characterize the spatial extent of Ly$\alpha$ emission, we fit the radial profile using two different models.

In the first model, we assume the \lyalpha\ surface brightness to be an exponentially declining function of projected galactocentric distance,
\begin{eqnarray}
\label{eq:expo}
S(r) = S_{\rm 0} \exp(-r/r_{\rm s}),
\end{eqnarray}
where $r_{\rm s}$ is the exponential scale-length, and $S_0$ is the surface brightness at the source center. 
This simple approach was adopted in several previous works \citep{Steidel:2011jk,Matsuda:2012fp,Feldmeier:2013fx,Momose:2014fe}, with the caveat that the center light profile could be heavily influenced by the PSF.
The measurements may also suffer from large uncertainties at large radii, where the surface brightness profile is lost in the background noise.
In our analysis, we only fit the radial profile in the range of $r=[1\farcs75,6\farcs00]$, similar to other analyses in the literature.
This radial range corresponds to a projected physical distance of 13\,--\,44\,kpc for the PCF sample, or of 14\,--\,49\,kpc for the LAB sample. 

The second model is motivated by \citet{Wisotzki:2016hw}, in which a \lyalpha\ radial profile is fit with two separate model components convolved with the image PSF: a core ``galaxy'' component with a compact exponential profile and a separate broader halo component which also declines exponentially.
The two-component model can be parameterized as
\begin{eqnarray}
\label{eq:twoc}
S(r) &=&  {\rm PSF} \ast \left[ S_{\rm c} \exp (-r/r_{\rm s, c})  + S_{\rm h}\exp(-r/r_{\rm s, h}) \right]\\
I(r) &=&   {\rm PSF} \ast I_{\rm c} \exp (-r/r_{\rm s, c})
\end{eqnarray}
where the exponential scales of two components are $r_{\rm s,h}$ and $r_{\rm s,c}$  (``c'' and ``h'' denote the continuum and halo components, respectively).
$S(r)$ and $I(r)$ present the \lyalpha\ and UV surface brightness, respectively. We simultaneously fit both of the \lyalpha\ and UV continuum using the above model, with $r_{\rm s, c}$ largely constrained by the continuum light profile.
This decomposition model automatically accounts for the PSF effect on both \lyalpha\ and UV continuum images, such that we can take a full advantage of the measured radial profile at all ranges. 
On the other hand, the single-component scale-length is expected to be sensitive to the selected projected galactocentric radii and large-scale PSF profile. We refer interested readers to  Appendix\,{\ref{sec:psf_effect}}, where we provide extensive discussions on this topic.

\begin{figure*}[!htbp]
\centering
\includegraphics[width=0.99\linewidth,angle=0]{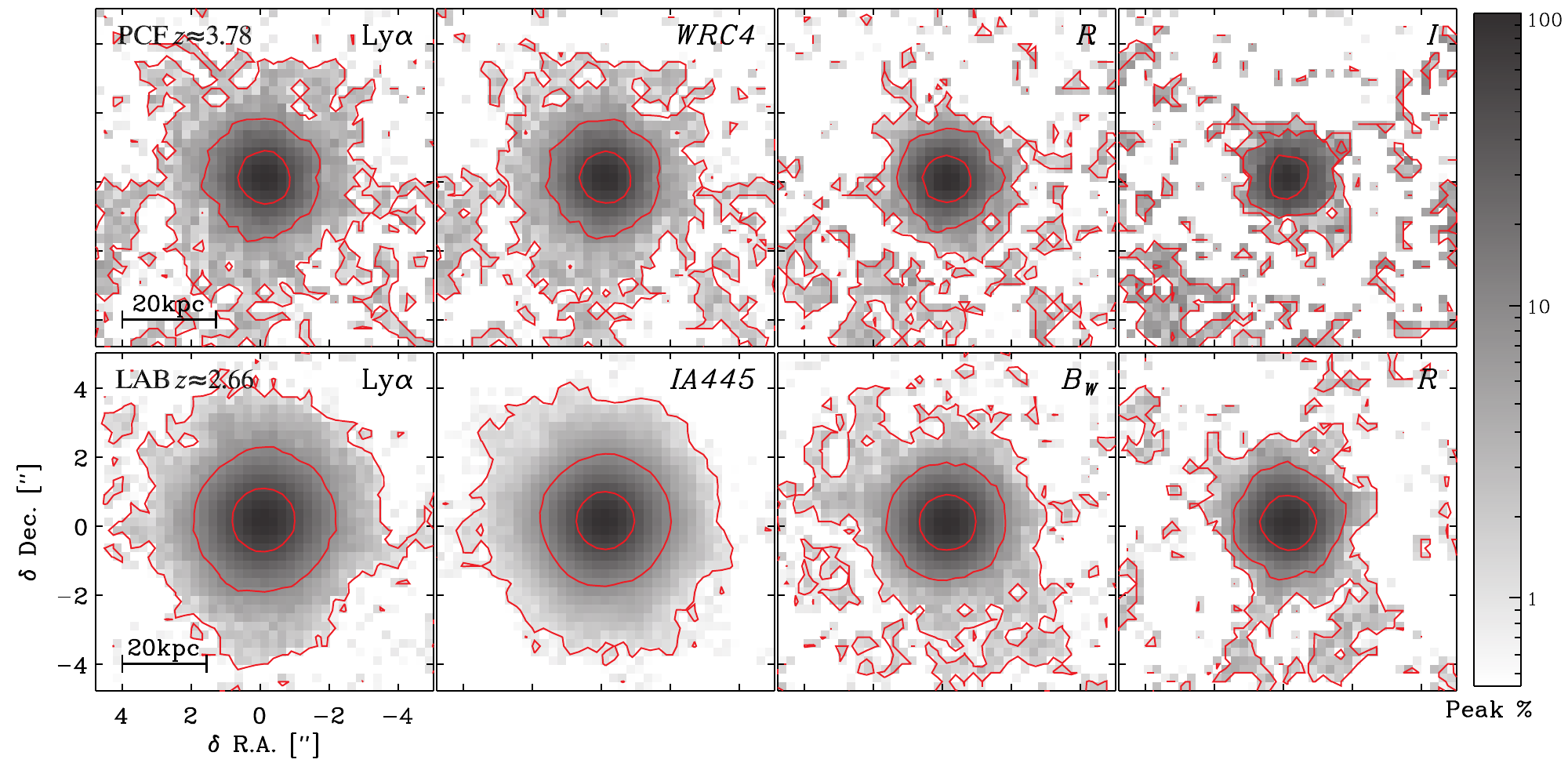}
\caption{\label{fig:radprofile_all_image}
Stacked images of the LAEs in the PCF (top) and LAB (bottom) fields. 
All images are produced by taking the pixel-to-pixel median of the resampled individual cutouts centered on the galaxy positions, and the Ly$\alpha$ images are derived from the difference between the narrow- and broadband data (see the descriptions in Section\,\ref{sec:defaultstack}). 
For the PCF (LAB) sample, the {\it WRC4} and $R$ ({\it IA445} and $B_W$) bands contain both of the Ly$\alpha$ and continuum emission near $\lambda_{\rm{rest}}\approx 1220$\,\AA, while the $I$ ($R$) band samples the UV continuum at $\lambda_{\rm{rest}}\approx 1700$\,\AA. 
For each image, the red contours show the positions at which the surface brightness falls to 50\%, 10\%, and 1\%  of the peak value. 
The peak-normalized intensity scale is indicated by the color bar on the right. Diffuse emission is present in the narrow and Ly$\alpha$ bands of both samples.
The angular distance scale is indicated at the bottom left corner of each panel.
}
\end{figure*}
\begin{figure*}[!htbp]
\includegraphics[width=1.0\linewidth]{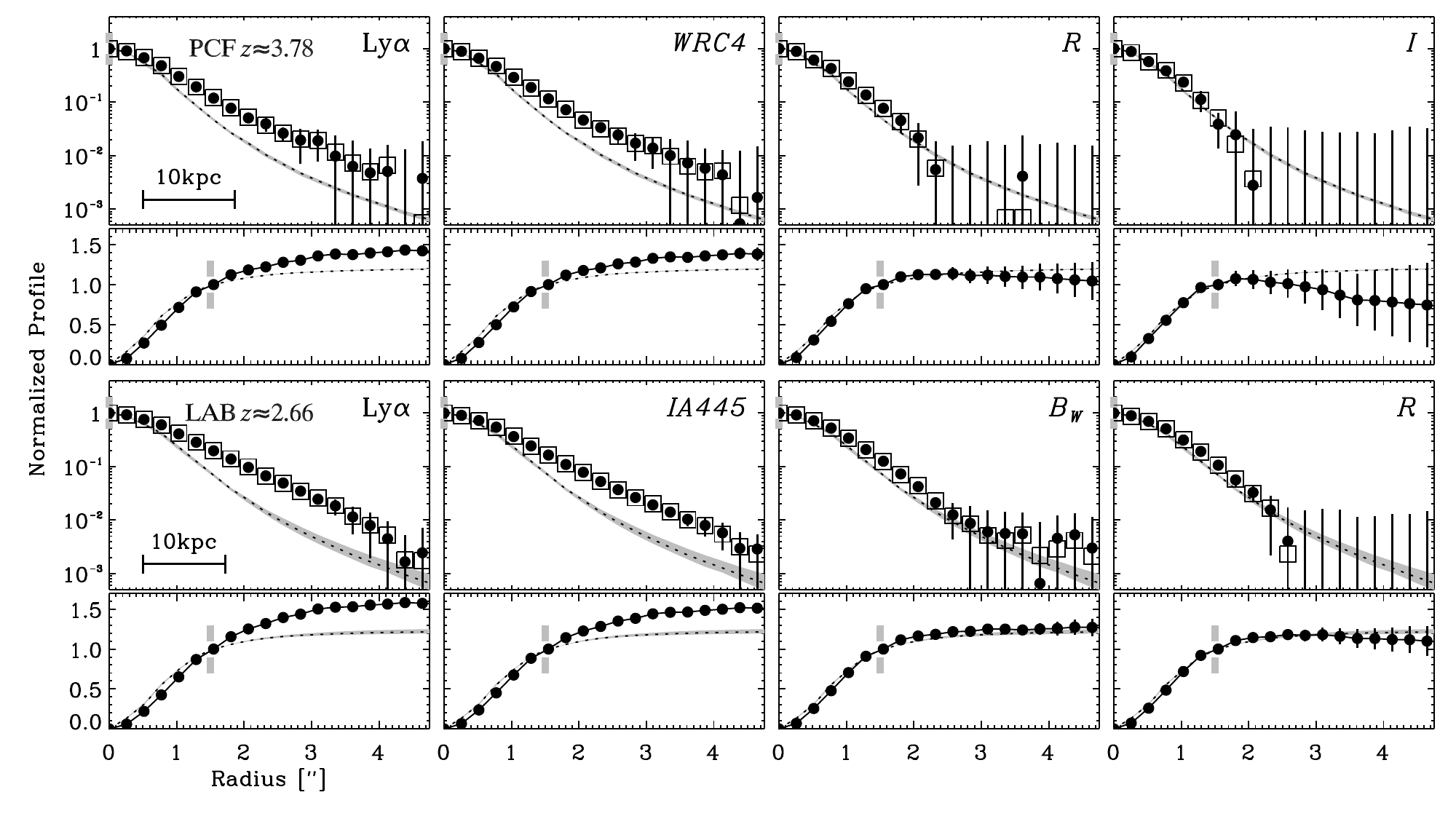}
\caption{\label{fig:radprofile_all_comp} 
Radial profiles of stacked images presented in Figure~\ref{fig:radprofile_all_image}. 
For each band, we show both differential and cumulative profiles, normalized at the center and $r=1\farcs5$ (indicated by the gray bars), respectively. The differential radial profile is measured by taking median (filled circles) and mean (open squares) in each annulus bin. 
For comparison, the normalized stellar PSFs are shown as dotted black lines. Gray shades indicate the variations of homogenized PSFs among different pointing and bands.
The angular distance scale is indicated at the bottom left corner of each panel.
}
\end{figure*}

\subsection{Definitive Detection of Diffuse Ly$\alpha$ Emission around Galaxies}

In Figures\,\ref{fig:radprofile_all_image} and \ref{fig:radprofile_all_comp}, we present the stacked galaxy images and their corresponding one-dimensional surface brightness radial profiles, all of which are constructed from the full LAE samples in the LAB and PCF fields at $z\approx2.66$ and $z\approx3.78$, respectively.
The emission in Ly$\alpha$ or the NB filters is clearly more extended than that in the broadbands or stellar PSFs, while the radial profiles of continuum band stacked images  ($I$ and $R$-band for the PCF and LAB sample, respectively) are consistent with their respective stellar PSFs.
We further test the validity of our detection by repeating the stacking procedure on a sample of randomly selected NB-detected sources within the same brightness range. The resulting radial profiles of stacked images are consistent with stellar PSFs in {\it all} bands, eliminating the possibility that the extended structures originate from large-scale PSF wings \citep{Feldmeier:2013fx} or artifacts from our stacking procedures. 

In Tables\,\ref{pcf_table} and \ref{lab_table}, we list the LAH scale-lengths estimated from two profile fitting methods described in Section\,\ref{sec:defaultstack}. 
Their uncertainties at 90\% confidence level are estimated using Monte Carlo simulations by repeatedly fitting models to new realizations of stacking light profiles with added systematic and statistical noise.
For both PCF ($z\approx3.78$) and LAB ($z\approx2.66$) samples, we measure the Ly$\alpha$ scale-lengths $r_{\rm s}$ to be $5$\,--\,$6$~kpc from the median stacked images. 
We note that the LAB sample does not include the known giant \lyalpha\ blobs discovered in \citet{Dey:2005dl} and \citet{Prescott:2012ed,Prescott:2013iu}.
However, their inclusion does not affect the stacking results due to their limited contributions in the full sample stacking.

\begin{figure}[!htbp]
\center
\includegraphics[width=0.40\paperwidth]{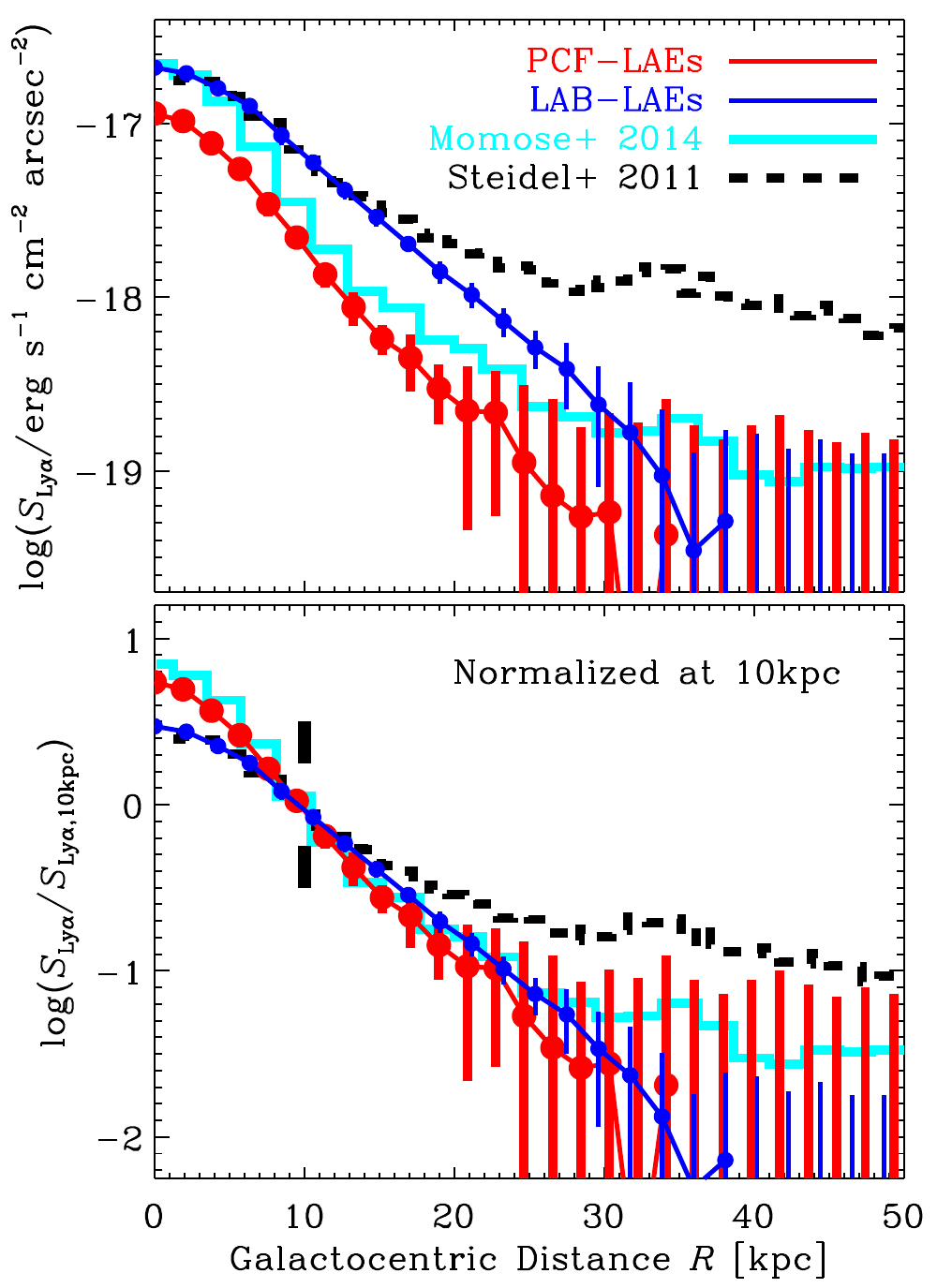}
\caption{\label{fig:radprofile_allobs}
Average \lyalpha\ radial profiles of our PCF-LAE (red) and LAB-LAE (blue) samples, together with similar measures in the literature: the $z=3.1$ LAE sample from \citet{Momose:2014fe} (cyan) and the ``LAE-only'' subsample from S11 (black). All measures are presented in their original forms, including varying degrees of blurring due to the respective image PSFs.
Top: the surface brightness levels are in physical units. 
Differences in the overall amplitudes reflect the sample variation in intrinsic Ly$\alpha$ luminosities and cosmological dimming.
Bottom: the same profiles are normalized at 10\,kpc to facilitate comparison of their overall profiles beyond the central galaxies (see the text). 
Both of our samples have profiles similar to that measured by \citet{Momose:2014fe} from 10 to 30\,kpc.
}
\end{figure}

Our size measurements are in excellent agreement with similar measurements performed on individual galaxies at $z\sim 3$\,--\,$6$  \citep{Wisotzki:2016hw} and with the median values found for $z\sim3$ LAEs \citep{Feldmeier:2013fx,Momose:2014fe}.  In contrast, the LAH exponential scale-lengths of our samples are factors of five to six smaller than that of the ``LAE only'' sample in S11. 
Their sample consists of 18 galaxies with ${\rm EW}_0({\rm Ly}\alpha)\geq 20$\,\AA.
In Figure~\ref{fig:radprofile_allobs}, we present the \lyalpha\ radial profiles measured from our LAE samples, together with previous results from S11 and \citet{Momose:2014fe}. 
The radial profile in the central region is heavily influenced by the significance of Ly$\alpha$ emission from the compact galaxies and the image PSF, and thus depends on the sample selection and the imaging data quality. 
Therefore, we normalize all profiles at 10\,kpc (or $\approx1\farcs3$\,--\,1\farcs4) in the bottom panel of Figure~\ref{fig:radprofile_allobs}, and focus on comparing the slope of the outer radial profile. 
This slope determines the exponential scale-length as given by Equation\,\ref{eq:expo}.  
As is evident in Figure~\ref{fig:radprofile_allobs}, our measured profiles agree reasonably with that of \citet{Momose:2014fe} out to 30\,kpc, while all of them are clearly at odds with the S11 profile, which declines much more slowly than the rest. 

The large discrepancy between our results and S11 may be explained if Ly$\alpha$ halo sizes depend on galaxy properties, such as UV continuum luminosities or Ly$\alpha$ luminosities.
The studies by \citet{Feldmeier:2013fx} and by \citet{Momose:2014fe} reported that the \lyalpha\ halos may be more prominent around galaxies with lower Ly$\alpha$ EWs and with brighter UV luminosities \citep[but see][]{Wisotzki:2016hw}.
While the \lyalpha\ EW distributions of our PCF sample and the S11 sample are comparable, the latter consists of continuum-detected LBGs with the median magnitude of $R=24.8$~AB, where the $R$-band traces the rest-frame UV continuum at $\sim1500$\,\AA.
In comparison, our PCF sample is detected in an NB filter, and color-selected to have high line EWs regardless of UV luminosity. 
The majority of the selected galaxies are actually continuum faint ($I\gtrsim 25.5$~AB).
Considering the cosmological dimming from the redshift difference and assuming the same UV continuum slope, the S11 sample should have, on average, both higher Ly$\alpha$ and UV luminosities than those of our PCF sample.
In practice, redder spectra are more likely in their sample because continuum bright galaxies tend to be redder \citep[e.g.][]{Bouwens:2009ik}. 
This will make the luminosity gap even larger.
Similarly, our LAB sample galaxies are even fainter in the absolute UV magnitude with a median of $M_{\rm UV}>-20$. 

The \lyalpha\ halo difference is also conceivably due to the halo size changing with local environments or evolving with redshift.
Notably, \citet{Matsuda:2012fp} analyzed LAHs in bins of LAE surface overdensities ($\Sigma_{\rm g}$), and reported that Ly$\alpha$ halo size scales as $r_{\rm s}\propto \Sigma_{\rm g}^2$. Their ``field'' LAEs have halo sizes of $8$\,--\,$11$\,kpc, considerably smaller than  $\sim20$\,kpc for the LAEs in the highest overdensity region. 
Such a strong dependence may also simultaneously explain the larger LAHs around the S11 sample galaxies, which exclusively reside in massive protoclusters.
\citet{Wisotzki:2016hw} measured the LAH sizes of individual LAEs in their sample, and reported that the LAHs around $z\sim3.7$ LAEs are a factor of approximately two larger than those around $z\sim5.1$ LAEs. However, their sample spans a wide range of Ly$\alpha$ EWs, UV and Ly$\alpha$ luminosities (and of an unknown range of galaxy overdensities), making it difficult to interpret whether the observed trend is a result of redshift evolution, or of varied physical properties in these LAEs at different redshifts.
Comparing our scale-length measurements of high-redshift LAEs with those of local Ly$\alpha$-emitting galaxies may also provide an insight to the redshift evolution scenario.
 \citet{Hayes:2013jc} measured the Petrosian radii\footnote{In \citet{Hayes:2013jc}, it is specifically defined as the radius at which the local surface brightness is 20\% of the average surface brightness inside.} \citep{Petrosian:1976ea}, $R_{\rm P20}^{\rm Ly\alpha}$, for 14 galaxies in their Ly$\alpha$ Reference Sample (LARS, hereafter), and found that they range from 3 to 15\,kpc.
By assuming that the intrinsic Ly$\alpha$ emission is symmetric and declines exponentially with projected galactocentric distance, we find that an exponential profile with a scale-length of 6\,kpc corresponds to $R_{\rm P20}^{\rm Ly\alpha}\approx22$~kpc.
Therefore, high-redshift LAEs in our sample appear to have more extended LAHs than the LARS galaxies.
However, we caution that our $R_{\rm P20}^{\rm Ly\alpha}$ estimates are only approximate and more likely upper limits. Properly accounting for the PSF blurring effect and adding a compact galaxy-like \lyalpha-emitting component (instead of a single exponential profile) may lead to smaller Petrosian radii. 

Despite the above speculations, the major reasons for the different halo sizes remain unclear.
We note that different scenarios may not be easy to disentangle from the aspect of sample selections. For example, galaxies residing in overdense regions may have enhanced Ly$\alpha$ or UV luminosities compared to those in underdense or field environments \citep[e.g.,][]{Koyama:2013ez, Lemaux:2014by, Dey:2016dl}. 
Analyses on large and well-controlled diverse galaxy samples within different enviroments will be crucial to further investigate the primary factor that determines LAH sizes, and thereby constrain the physical origin of LAHs.
This is the subject of Sections \ref{sec:measure_overdensity} and \ref{sec:results}.

\section{Characterizing Galaxy Environments}\label{sec:measure_overdensity}

Galaxies in our sample reside in a wide range of environments, including at least three massive protoclusters \citep[][also Section\,\ref{sec:datasample}]{Dey:2005dl,Prescott:2012kp,Lee:2014gv,Dey:2016dl}. 
Hence, we are well positioned to explore how the LAH characteristics depend on local galaxy density. 
We begin by estimating the local LAE overdensity for each sample. 

\subsection{Estimation of LAE Overdensities}
\label{sec:laeoverd}

For the PCF sample, the NB filter width corresponds to an effective line-of-sight  comoving distance of 27\,Mpc (or 5.6\,Mpc in physical distance), qualitatively comparable to the expected size of (unvirialized) forming clusters at the same redshift \citep{Chiang:2013fs}. 
The redshift distribution of PC~217.96+32.3 peaks sharply close to the value corresponding to the central wavelength of the NB filter \citep{Dey:2016dl}.
Hence, the use of LAE surface overdensity with smoothing scales of ${\rm FWHM}=5$\,--\,10\,Mpc is well justified as a proxy for the three-dimensional overdensity with minimal contamination by foreground and background interlopers. 

We determine the surface overdensity, $\Sigma_{\rm LAE}$, as a function of image positions as follows. First, the two-dimensional LAE distribution map is smoothed with a Gaussian kernel of ${\rm FWHM}=10$\,Mpc (or $\sigma=2\arcmin$). Second, the field LAE density, $\overline{\Sigma}_{\rm LAE}$, is estimated by dividing the number of LAEs within the effective area (excluding masked regions) after removing the sources likely associated with PC~217.96+32.3.
Finally, the LAE surface overdensity is computed as
\begin{equation}
\label{eq:sdelta}
\delta_{\rm LAE}=\frac{\Sigma_{\rm LAE}-\overline{\Sigma}_{\rm LAE}}{\overline{\Sigma}_{\rm LAE}}.
\end{equation}
While all galaxies, including LBGs, are assigned an overdensity parameter based on their image positions, the overdensity itself is determined solely based on the angular distribution of LAEs. 

As for the LAB sample at $z\approx2.66$, the same technique is unlikely to yield galaxy overdensity parameters in a meaningful manner. 
The intermediate filter {\it IA445}, used to identify photometric LAE candidates at $z\approx2.66$,  spans a line-of-sight distance of 190\,Mpc. 
The shot noise from interlopers is expected to play a more significant role in the measured surface overdensity. 
Therefore, we decided to limit our density-related analyses to the spectroscopic sample only, and estimate galaxy overdensity in a three-dimensional comoving volume.
First, we construct a 3D LAE map based on their image positions and redshift, assuming the Ly$\alpha$ redshift as systemic redshift. 
Second, the map is smoothed with a 3D Gaussian kernel of ${\rm FWHM}=20$\,Mpc.
Finally, the 3D overdensity, $\delta_{\rm LAE}$, is computed using the smoothed LAE volume density.
As described in Section\,\ref{lab_data}, our Hectospec observations resulted in a relatively uniform spectroscopic coverage only in the middle two thirds of the field. 
Because of the potential bias in estimating the LAE overdensity in the eastern and western end of the field (totaling $\sim 0.3$~deg$^2$ in area), we excluded any spectroscopic candidates there from our subsequent analysis related to the overdensity property.
The use of Ly$\alpha$ redshift in determination of galaxy line-of-sight position can also lead to an error in our overdensity estimation.
However, such uncertainties are likely negligible here. 
A typical offset between Ly$\alpha$-derived and systemic redshift is $\Delta z=0.003$\,--\,$0.004$ or $250$\,--\,$330$~\kms\ in velocity space, which will lead to positional errors of 3.4\,--\,4.6\,Mpc comoving, much smaller than the adopted density smoothing scale of 20\,Mpc.
Realistically, the positional error is usually smaller than the above values because most LAEs have their Ly$\alpha$ line shifted in the same direction with similar magnitudes.

In Figure\,\ref{fig:pcflab_env}, we show the overdensity properties of both the PCF field (left) and the LAB field (right), together with the positions of individual sources in each sample, color coded in redshift for the spectroscopic sources. 
For the PCF field, both contours and orange shades indicate the surface overdensity levels, clearly marking two spectroscopically confirmed structures located at the center and the northeastern corner of the PCF field \citep{Lee:2014gv, Dey:2016dl}. 
LBG members of the structures are marked as diamonds, though they are not included in the overdensity estimate. 
In the LAB field, we show the second moment of the three-dimensional overdensities as orange shades. The highest overdensity region ($\delta_{\rm LAE}\gtrsim 5$) is located slightly off center in the southeastern direction, coinciding with the position of the LAB, LABd05, marked as a large open square \citep{Dey:2005dl}. 

Our choices of smoothing kernels (10 and 20\,Mpc for the PCF and LAB samples, respectively) are justified by the observed surface density of LAEs in these fields. 
A smoothing kernel size -- i.e., the effective volume within which the galaxies are counted -- should be large enough to enclose a sufficient number of galaxies to minimize shot noise in the density estimate. 
The transverse distance to the nearest neighbor for the PCF sample ranges from 0.3 to 9.7\,Mpc with a median value of 2.7\,Mpc. 
As for the LAB sample, the total (transverse) distance to the nearest neighbor spans 0.7\,(0.4)\,Mpc to 31.5\,(18.1)\,Mpc with the median of 7.8\,(4.5)\,Mpc. Therefore, the adopted kernels are large enough to enclose at least one LAE in the least concentrated regions.

\subsection{Calibration of Local Overdensities}

The above overdensity estimates should be a good relative measure to distinguish overdense or underdense regions within a single field. 
However, they are inadequate to be compared with one another or with similar measures in other surveys, because the overdensity parameter is a strong function of imaging or spectroscopic depth of the survey, filter-determined redshift range, and smoothing scale at which it is derived \citep[e.g.][]{Chiang:2013fs}.
The imaging and spectroscopic depth not only influence the overall error budget due to shot noise but also determine the average bias of the galaxy population uncovered as a result of such a survey.
The NB filter width would decide the shot noise from unassociated sources in the foreground and background, which is mixed with the signal from coherent structures.
A large smoothing kernel (2D or 3D) would effectively average over more galaxies, reducing the shot noise of the overdensity measure at the expense of washing out features that are smaller than the size of the kernel. 
Conversely, using too small a kernel would exaggerate the overdensity parameter.
Depending on the specifics of a galaxy density structure and how it is observed, a range of overdensity parameters is expected for the same structure. 

To facilitate direct comparison of our PCF and LAB results with other measurements in the literature, we calibrate our overdensity estimates using a simulated $z\sim3$ galaxy catalog of \citet{Chiang:2013fs}. 
This catalog is based on the Millennium I and II Simulation Runs \citep{Springel:2005gv}, and provides sky positions and redshift of individual galaxies with star-formation rates $\geq 1M_\odot$\,yr$^{-1}$, comparable to typical values measured for LAEs \citep[e.g.,][]{Gawiser:2006fv, Guaita:2010kr}. 
Specifically, we first derive galaxy overdensities in the simulated comoving volume using a top hat cubic box with 15\,Mpc on a side. Second, each simulated galaxy is assigned an overdensity parameter accordingly. 
Then, we repeated the steps \#1 and \#2 using a series of different smoothing kernels closely matching those used in our data and also those used by \citet{Matsuda:2012fp} and \citet{Momose:2016cu} (see Table~\ref{tab:pworks}).
We denote the respective overdensity parameters as $\delta^{\rm PCF}_{\rm sim}$, $\delta^{\rm LAB}_{\rm sim}$, $\delta^{\rm M12}_{\rm sim}$, and $\delta^{\rm M16}_{\rm sim}$.
Our simulation results from the same set of galaxies determine how these measures map into the overdensity measured in a 15\,Mpc cubic box, or $\delta^{\rm 15\,Mpc}_{\rm sim}$, which is adopted as the benchmark to evaluate the significance of a protocluster. 

\begin{figure}[t]
\centering
\includegraphics[width=0.40\paperwidth]{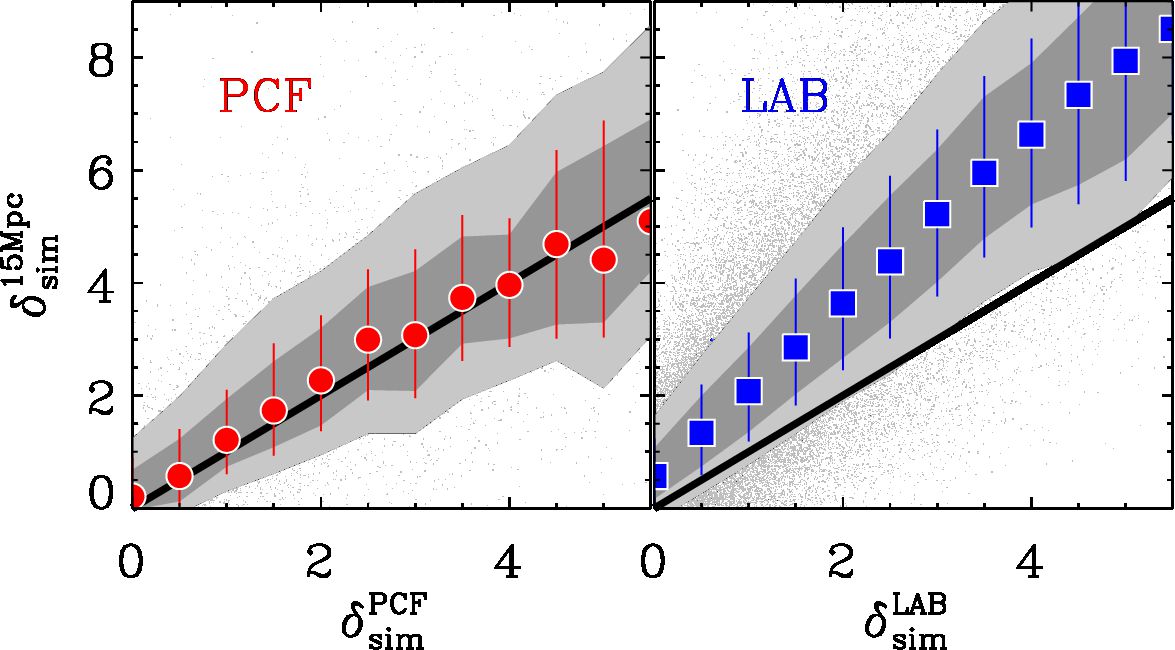}
\caption{\label{fig:deltacomp}
Calibration of measured galaxy overdensities: 
utilizing a galaxy catalog from the Millennium Runs, we determine local galaxy overdensities using smoothing kernels matching those used in the real data, and compare them to the values determined with a (15\,Mpc)$^3$ top hat cubic kernel, $\delta_{\rm{sim}}^{\rm{15Mpc}}$.
Dark and light gray shades bracket the 30\,--\,70\% and 15\,--\,85\% range of $\delta_{\rm{sim}}^{\rm{15Mpc}}$ scatter at a specific overdensity based on the kernel applied to the observational data, respectively.
In both panels, the thick black line marks the one-to-one line. The correction factor derived from the median relation is later used to bring all measured overdensities in real data to a common standard in $\delta_{\rm{sim}}^{\rm{15Mpc}}$.
}
\end{figure}

In Figure~\ref{fig:deltacomp}, we show the distribution of overdensity parameter of all simulated galaxies on the $\delta^{\rm 15\,Mpc}_{\rm sim}$--$\delta^{\rm PCF}_{\rm sim}$ (left) and  $\delta^{\rm 15\,Mpc}_{\rm sim}$--$\delta^{\rm LAB}_{\rm sim}$ plane (right). The $\delta^{\rm PCF}_{\rm sim}$ parameter is close to our benchmark: the larger line-of-sight distance (25~Mpc) and  smaller kernel size (${\rm FWHM}=10$\,Mpc) together effectively enclose a cosmic volume similar to our benchmark measure.  On the other hand, the $\delta^{\rm LAB}_{\rm sim}$ parameter yields a systematically lower value than the benchmark mainly due to the larger smoothing scale (${\rm FWHM}=20$\,Mpc). 
The same procedure is carried out for the \citet{Matsuda:2012fp} and \citet{Momose:2016cu} estimates, and all overdensity parameters measured from real data are converted to the benchmark values $\delta^{\rm 15\,Mpc}_{\rm sim}$.

\section{Dependence of LAH\MakeLowercase{s} on Galaxy Properties}
\label{sec:results}

We now investigate how the LAH properties depend on local environment, Ly$\alpha$ luminosities and equivalent widths, and UV continuum luminosities. 
The LAE overdensities (detailed in Section\,\ref{sec:measure_overdensity}) are used to characterize the environment of individual galaxies.
Because all of our samples are LAEs (except the small number of LBGs in the PCF field) and confined into narrow redshift ranges\footnote{The luminosity estimation should be robust even for the non-spectroscopic LAEs in the LAB field ($z=2.569$\,--\,$2.737$) because the redshift uncertainty only leads to a brightness error of $\approx0.11$\,mag.}, their UV luminosities, Ly$\alpha$ luminosities, and line EWs can be ranked by their CB brightness, NB magnitudes, and NB$-$BB colors, respectively.
Based on these directly measured parameters, we subdivide our samples accordingly and repeat the image stacking and profile measurement procedures as described in Section\,\ref{sec:defaultstack}.

\begin{figure*}[!htbp]
\begin{minipage}{0.49\linewidth}
\includegraphics[width=1.0\linewidth]{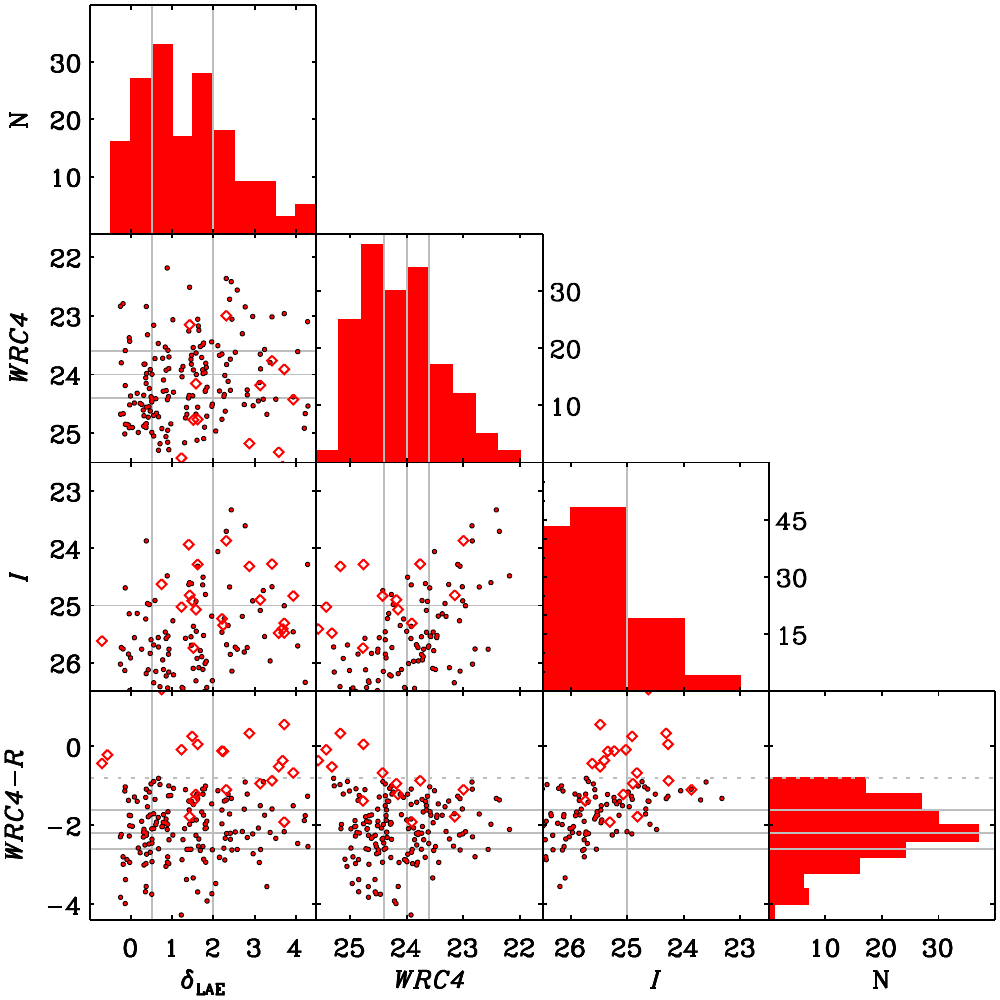}
\end{minipage}
\begin{minipage}{0.49\linewidth}
\includegraphics[width=1.0\linewidth]{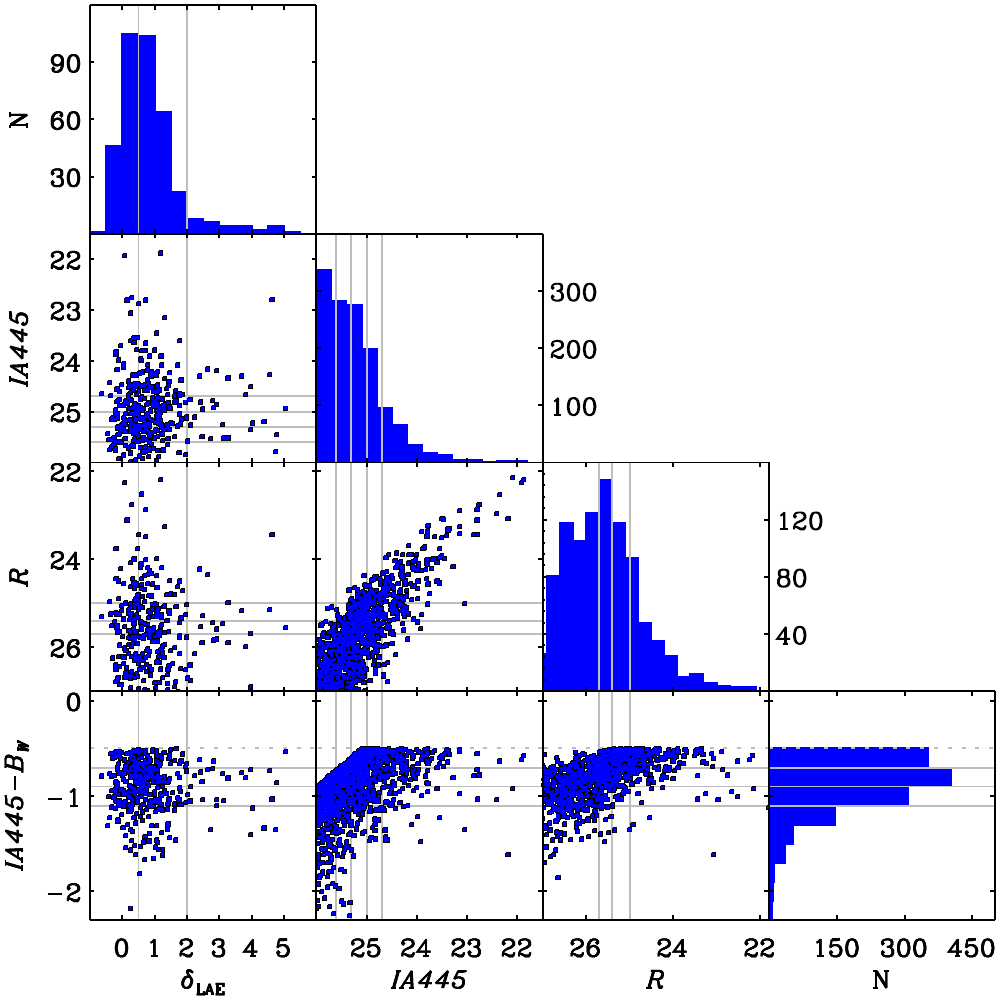}
\end{minipage}
\caption{\label{fig:pcflab_propc}
Distributions of measured galaxy properties within our PCF (red) and LAB (blue) LAE samples: namely,  LAE overdensities, NB/CB magnitudes, and NB$-$BB colors. For the PCF sample, the properties of the spectroscopically confirmed LBGs at the same redshift are indicated as open diamonds. Solid gray lines represent the divide between subsamples as described in Section\,\ref{sec:results}. 
In both samples, a positive correlation exists between NB and CB magnitudes, which is a bias from LAE color selection criteria \citep[see, e.g.][]{Ciardullo:2012gt}.
}
\end{figure*}

In Tables~\ref{pcf_table} and \ref{lab_table}, we list the criteria and size of each subsample in the PCF and LAB field, respectively.
The distributions of galaxy properties and subsample boundaries are presented in Figure\,\ref{fig:pcflab_propc}.
Also shown in the tables are the best-fit scale-lengths using both the exponential model and the two-component model (see Equations~\ref{eq:expo} and \ref{eq:twoc} in Section\,\ref{sec:defaultstack}). 
The exponential scale-length, $r_{\rm s}$, does not explicitly account for the PSF, and thus can be sensitive to the exact PSF shape at outer radii. A more extensive discussion on this subject is given in Appendix\,\ref{sec:psf_effect}. 

We also present the average physical properties of each subsample in Tables~\ref{pcf_table} and \ref{lab_table}: Ly$\alpha$ luminosity, equivalent width, and UV luminosity.
The absolute UV magnitudes are computed from the broadband brightness near the rest-frame wavelength 1700\,\AA\ (assuming $f_\lambda \propto \lambda^{-2}$), over an aperture of radius $r=2\arcsec$.
The Ly$\alpha$ luminosities and rest-frame EWs within an $r=3\arcsec$ aperture are derived fully taking into account the filter response and intergalactic absorption, as detailed in Appendix\,\ref{sec:flya}.
The slightly larger aperture is chosen to capture the total Ly$\alpha$ flux of stacked objects (see the cumulative \lyalpha\ light profile in Figure~\ref{fig:radprofile_all_comp}).
These values are derived directly from the stacked images of individual subsamples, but agree with the median of the same quantities determined individually from each galaxy.

\floattable
\begin{deluxetable}{ccccccccch}[htb!]
\tabletypesize{\footnotesize}
\tablecolumns{7} 
\tablecaption{PCF Subsamples and Their Average Continuum and LAH Properties\label{pcf_table}}
\tablewidth{0pt} 
\tablehead{
\multicolumn{1}{c}{Name}&
\multicolumn{1}{c}{Selection}&
\multicolumn{1}{c}{$N$\tablenotemark{a}}&
\multicolumn{1}{c}{$\log(L_{\rm Ly\alpha})$\tablenotemark{b}}&
\multicolumn{1}{c}{$M_{\rm UV}$\tablenotemark{c}}&
\multicolumn{1}{c}{EW$_{0}$\tablenotemark{d}}&
\multicolumn{1}{c}{$r_{\rm s}$}&
\multicolumn{1}{c}{$r_{\rm s,h}$}&
\multicolumn{1}{c}{$r_{\rm s,c}$\tablenotemark{e}}\\
\multicolumn{1}{c}{}&
\multicolumn{1}{c}{}&
\multicolumn{1}{c}{}&
\multicolumn{1}{c}{(erg s$^{-1}$)}&
\multicolumn{1}{c}{}&
\multicolumn{1}{c}{\AA}&
\multicolumn{1}{c}{(kpc)}&
\multicolumn{1}{c}{(kpc)}&
\multicolumn{1}{c}{(kpc)}
} 
\startdata
LBG & Equation\,\ref{eq:cc-lbg} (Section\,\ref{pcf_data}) & 21 & $42.42$ & $-21.24^{+0.17}_{-0.15}$ & $10^{+5}_{-3}$ & $7.5^{+6.3}_{-3.7}$ & $5.8^{+5.2}_{-1.3}$ & 1.3\\
LAE & Equation\,\ref{eq:cc-lae} (Section\,\ref{pcf_data}) & 163 & $42.72$ & $-19.79^{+0.14}_{-0.13}$ & $70^{+14}_{-11}$ & $5.7^{+0.7}_{-0.5}$ & $4.8^{+0.4}_{-0.9}$ & 1.2 \\
\midrule
LAE-env-s1 & $2.0\leq\delta_{\rm LAE}$ & 44 & $42.87$ & $-20.32^{+0.20}_{-0.16}$ & $67^{+12}_{-10}$ & $5.5^{+1.0}_{-0.8}$ & $5.2^{+1.4}_{-0.9}$ & 1.7 \\
LAE-env-s2 & $0.5\leq\delta_{\rm LAE}<2.0$  & 76 & $42.75$ & $-19.47^{+0.25}_{-0.21}$ & $87^{+14}_{-11}$ & $7.2^{+1.4}_{-1.1}$ & $7.0^{+2.1}_{-2.5}$ & 1.4  \\
LAE-env-s3 & $\delta_{\rm LAE}<0.5$  & 43 & $42.72$ & $-19.24^{+0.45}_{-0.25}$ & $100^{+40}_{-24}$ & $4.2^{+2.3}_{-0.7}$ & $3.9^{-1.9}_{-0.6}$ & 0.9 \\
\midrule
LAE-\lyalpha-s1& ${\it WRC4}<23.6$  & 36 & $43.08$ & $-20.81^{+0.15}_{-0.11}$ & $67^{+12}_{-9}$ & $7.2^{+1.2}_{-1.0}$ & $7.0^{+1.5}_{-1.1}$ & 1.5 \\
LAE-\lyalpha-s2& $23.6\leq {\it WRC4}<24.0$  & 34 & $42.84$ & $-19.18^{+0.27}_{-0.22}$ & $113^{+35}_{-28}$ & $7.4^{+1.1}_{-0.9}$ & $8.6^{+3.7}_{-2.9}$ & 1.5 \\
LAE-\lyalpha-s3& $24.0\leq {\it WRC4}<24.4 $  & 29 & $42.74$ & $-19.79^{+0.48}_{-0.33}$ & $68^{+21}_{-14}$ & $4.7^{+2.2}_{-1.4}$ & $3.5^{+2.1}_{-1.3}$ & 1.1 \\
LAE-\lyalpha-s4& $24.4\leq {\it WRC4} $  & 64 & $42.57$ & $-19.06^{+0.59}_{-0.38}$ & $73^{+39}_{-22}$ & $3.8^{+0.5}_{-1.1}$ & $2.5^{+0.4}_{-0.7}$ & 0.9  \\
\midrule
ALL-UV-s1& $I<25.0$  & 19 & $42.88$ & $-21.58^{+0.07}_{-0.07}$ & $20^{+5}_{-4}$ & $6.9^{+1.2}_{-1.5}$ & $5.5^{+2.7}_{-0.6}$ & 1.3  \\
ALL-UV-s2& $25.0 \leq I$  & 161 & $42.74$ & $-19.71^{+0.14}_{-0.12}$ & $72^{+11}_{-9}$ & $5.4^{+1.1}_{-0.4}$ & $3.7^{+0.9}_{-0.5}$ & 1.1  \\
\midrule
LAE-EW-s1 & ${\it WRC4}-R <-2.6$ & 44 & $42.74$ & $-17.65^{+0.89}_{-0.55}$ & $242^{+370}_{-165}$ & $3.9^{+0.5}_{-0.4}$ & $2.8^{+0.3}_{-1.2}$ & 1.0   \\
LAE-EW-s2 & $-2.6 \leq {\it WRC4}-R <-2.2$ & 28 & $42.80$ & $-19.20^{+0.52}_{-0.42}$ & $136^{+74}_{-36}$ & $5.9^{+2.9}_{-1.8}$ & $4.2^{+2.4}_{-0.8}$ & 0.9  \\
LAE-EW-s3 & $-2.2 \leq {\it WRC4}-R <-1.6$ & 47 & $42.82$ & $-19.93^{+0.20}_{-0.15}$ & $70^{+9}_{-7}$ & $8.2^{+1.2}_{-1.1}$ & $9.9^{+2.3}_{-2.4}$ & 1.8  \\
LAE-EW-s4 & $-1.6\leq {\it WRC4}-R $ & 44 & $42.76$ & $-20.85^{+0.17}_{-0.16}$ & $32^{+5}_{-3}$ & $6.2^{+1.2}_{-1.0}$ & $4.4^{+1.0}_{-1.1}$ & 1.4\\
\enddata
\tablenotetext{a}{The total number of galaxies considered in stacking analyses.}
\tablenotetext{b}{Average \lyalpha\ luminosity integrated over an $r=3\arcsec$ aperture in the stacked image. The typical uncertainty is $\sim$0.02\,dex base on the image statistical noise. However, the uncertainty in the LBG subsample reaches 0.1\,dex.}
\tablenotetext{c}{Absolute UV magnitude derived from the integrated flux in the stacked $I$-band image, over an $r=2\arcsec$ aperture.
The error represents the statistical uncertainty in the stacked image.}
\tablenotetext{d}{Rest-frame photometric EW derived from the ${\it WRC4}-R$ color of the stacked object, measured in an $r=3\arcsec$ aperture.
The error represents the uncertainty propagated from the stacked image noise.}
\tablenotetext{e}{All UV continuum light profiles are consistent with the expectation from a point source. The pixel size of 0\farcs258 is equivalent with an angular size of 1.9\,kpc at $z=3.78$.}
\end{deluxetable}
\floattable
\begin{deluxetable}{ccccccccch}[htb!]
\tabletypesize{\footnotesize}
\tablecolumns{7} 
\tablecaption{LAB Subsamples and Their Average Continuum and LAH Properties\label{lab_table}}
\tablewidth{0pt} 
\tablehead{
\multicolumn{1}{c}{Name}&
\multicolumn{1}{c}{Selection}&
\multicolumn{1}{c}{$N$\tablenotemark{a}}&
\multicolumn{1}{c}{$\log(L_{\rm{Ly}\alpha})$\tablenotemark{b}}&
\multicolumn{1}{c}{$M_{\rm UV}$\tablenotemark{c}}&
\multicolumn{1}{c}{EW$_{0}$\tablenotemark{d}}&
\multicolumn{1}{c}{$r_{\rm s}$}&
\multicolumn{1}{c}{$r_{\rm s,h}$}&
\multicolumn{1}{c}{$r_{\rm s,c}$\tablenotemark{e}}\\
\multicolumn{1}{c}{}&
\multicolumn{1}{c}{}&
\multicolumn{1}{c}{}&
\multicolumn{1}{c}{(erg s$^{-1}$)}&
\multicolumn{1}{c}{}&
\multicolumn{1}{c}{\AA}&
\multicolumn{1}{c}{(kpc)}&
\multicolumn{1}{c}{(kpc)}&
\multicolumn{1}{c}{(kpc)}
} 
\startdata
LAE  & Equation\,\ref{eq:cc-lab-lae} (Section\,\ref{lab_data})&  1336 & $42.83$ & $-19.20^{+0.08}_{-0.08}$ & $150^{+15}_{-13}$ & $6.0^{+0.5}_{-0.5}$ & $5.7^{+0.6}_{-0.5}$ & 1.7  \\ 
LAE-specz  & Equation\,\ref{eq:cc-lab-lae} (Section\,\ref{lab_data}) &  429& $42.89$ & $-19.36^{+0.05}_{-0.05}$ & $160^{+24}_{-20}$ & $5.5^{+0.5}_{-0.4}$ & $5.1^{+0.5}_{-0.5}$ & 1.2 \\ \midrule
LAE-specz-env-s1 & $2.0\leq\delta$ & 29 & $42.90$ & $-19.14^{+0.24}_{-0.21}$ & $177^{+57}_{-39}$ & $4.8^{+1.3}_{-0.9}$ & $5.1^{+1.7}_{-1.1}$ & 0.9 \\
LAE-specz-env-s2 & $0.5 \leq \delta<2.0$ & 139 & $42.92$ & $-19.32^{+0.09}_{-0.10}$ & $153^{+41}_{-23}$ & $5.6^{+0.6}_{-0.5}$ & $5.5^{+0.7}_{-0.5}$ & 1.5  \\
LAE-specz-env-s3 & $\delta<0.5$ & 86 & $42.94$ & $-19.30^{+0.14}_{-0.11}$ & $176^{+29}_{-21}$ & $5.7^{+0.7}_{-1.6}$ & $5.0^{+0.7}_{-0.6}$ & 1.0  \\
\midrule
LAE-\lyalpha-s1& ${\it IA445}<24.7$  & 228 & $43.19$ & $-20.51^{+0.03}_{-0.02}$ & $99^{+9}_{-7}$ & $6.5^{+0.7}_{-0.6}$ & $7.6^{+1.1}_{-1.0}$ & 2.0 \\
LAE-\lyalpha-s2& $24.7 \leq {\it IA445}<25.0$  & 162 & $42.94$ & $-19.78^{+0.07}_{-0.06}$ & $114^{+19}_{-12}$ & $6.3^{+0.7}_{-0.6}$ & $6.0^{+0.9}_{-0.7}$ & 1.8   \\
LAE-\lyalpha-s3& $25.0 \leq {\it IA445}<25.3$  & 289 & $42.82$ & $-19.47^{+0.06}_{-0.06}$ & $133^{+25}_{-16}$ & $6.1^{+0.5}_{-0.5}$ & $5.2^{+0.5}_{-0.4}$ & 1.5  \\
LAE-\lyalpha-s4& $25.3 \leq {\it IA445}<25.6$  & 278 & $42.74$ & $-18.58^{+0.10}_{-0.09}$ & $161^{+27}_{-23}$ & $5.6^{+1.1}_{-0.8}$ & $4.5^{+0.9}_{-0.7}$ & 0.7   \\
LAE-\lyalpha-s5& $25.6<{\it IA445}$  & 439 & $42.60$ & $-17.71^{+0.28}_{-0.22}$ & $216^{+34}_{-30}$ & $5.0^{+0.8}_{-0.6}$ & $3.9^{+0.6}_{-0.4}$ & 0.6   \\
\midrule
LAE-UV-s1& $R<25.0$  & 215 & $43.14$ & $-20.71^{+0.03}_{-0.02}$ & $74^{+11}_{-7}$ & $7.1^{+0.9}_{-0.7}$ & $8.6^{+1.2}_{-1.3}$ & 2.2  \\
LAE-UV-s2& $25.0 \leq R<25.4$  & 162   & $42.83$ & $-19.93^{+0.07}_{-0.05}$ & $90^{+12}_{-10}$ & $6.4^{+0.7}_{-0.6}$ & $5.5^{+0.7}_{-0.6}$ & 1.8 \\
LAE-UV-s3& $25.4 \leq R<25.7$  & 152 & $42.82$ & $-19.60^{+0.10}_{-0.08}$ & $101^{+17}_{-13}$ & $5.9^{+1.0}_{-0.9}$ & $4.7^{+1.5}_{-0.9}$ & 1.8  \\
LAE-UV-s4& $25.7 \leq R $  & 723 & $42.72$ & $-18.52^{+0.09}_{-0.08}$ & $228^{+52}_{-44}$ & $5.4^{+0.5}_{-0.4}$ & $4.2^{+0.9}_{-0.7}$ & 0.6 \\
\midrule
LAE-EW-s1 & ${\it IA445}-B_W<-1.1$ & 293 & $42.75$ & $-17.42^{+0.34}_{-0.21}$ & $279^{+125}_{-110}$ & $5.7^{+1.7}_{-1.2}$ & $4.6^{+2.1}_{-1.1}$ & 1.0     \\
LAE-EW-s2 & $-1.1\leq {\it IA445}-B_W<-0.9$ & 317  & $42.77$ & $-18.60^{+0.10}_{-0.07}$ & $206^{+40}_{-34}$ & $6.2^{+1.1}_{-0.9}$ & $5.2^{+1.2}_{-0.8}$ & 1.3  \\
LAE-EW-s3 & $-0.9\leq {\it IA445}-B_W<-0.7$ & 424  & $42.80$ & $-19.28^{+0.06}_{-0.05}$ & $127^{+16}_{-13}$ & $5.4^{+0.5}_{-0.4}$ & $4.9^{+0.5}_{-0.4}$ & 1.5   \\
LAE-EW-s4 & $-0.7\leq {\it IA445}-B_W$ & 362 & $42.90$ & $-20.11^{+0.04}_{-0.03}$ & $78^{+12}_{-17}$ & $6.8^{+0.7}_{-0.5}$ & $6.9^{+0.8}_{-0.5}$ & 2.0\\
\enddata
\tablenotetext{a}{The total number of galaxies considered in stacking analyses.}
\tablenotetext{b}{Average \lyalpha\ luminosity integrated over an $r=3\arcsec$ aperture in the stacked image. The typical uncertainty is $\sim$0.02\,dex base on the image statistical noise.}
\tablenotetext{c}{Absolute UV magnitude derived from the integrated flux in the stacked $R$-band image, over an $r=2\arcsec$ aperture.}
\tablenotetext{d}{Rest-frame photometric EW derived from the ${\it IA445}-B_W$ color of the stacked object, measured in an $r=3\arcsec$ aperture.}
\tablenotetext{e}{All UV continuum light profiles are consistent with the expectation from a point source. The pixel size of 0\farcs258 is equivalent with an angular size of 2.1\,kpc at $z=2.66$.}
\end{deluxetable}

\begin{figure}\centering
\includegraphics[width=0.40\paperwidth]{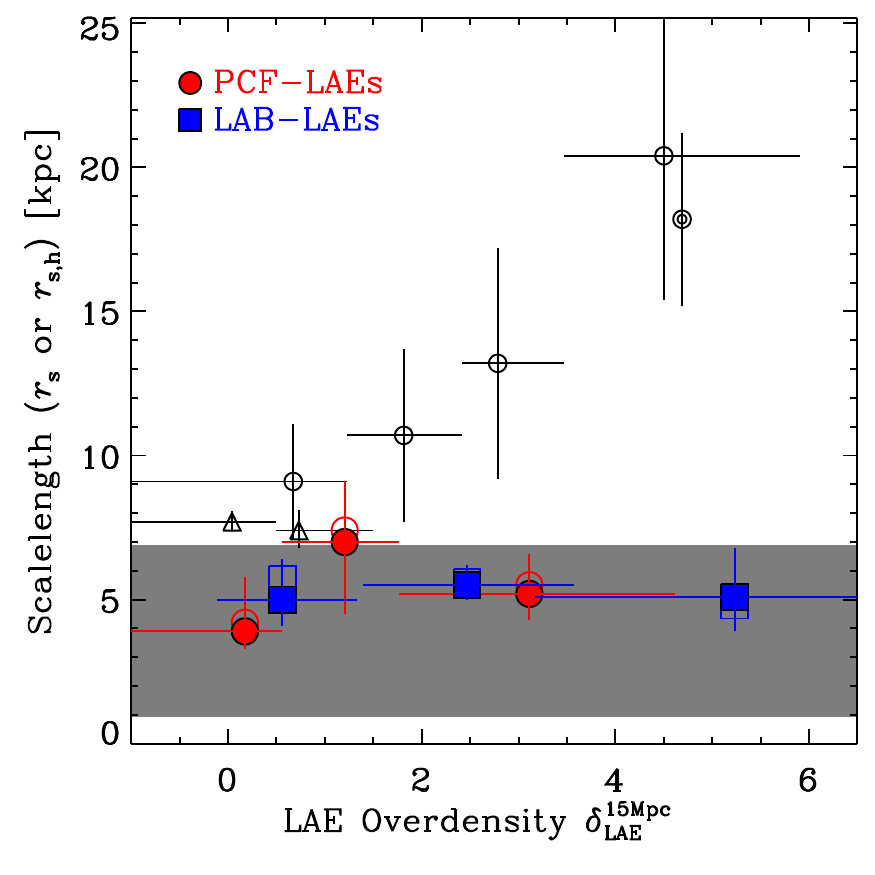}
\caption{\label{fig:slen_env}
Best-fit LAH scale-lengths as a function of LAE overdensity. 
The scale-lengths are measured by fitting surface brightness profiles to a single-component exponential model or a two-component model, denoted as $r_{\rm s}$ (open symbols) or $r_{\rm s,h}$ (filled symbols).
Our measurements are presented in red (PCF) and blue (LAB) symbols, and the results of \citet{Matsuda:2012fp} and \citet{Momose:2016cu} are shown in black open circles and triangles, respectively.
The gray shading indicates the LAH scale-length range measured  from individual galaxies in \citet{Wisotzki:2016hw}.
We note that \citet{Matsuda:2012fp} and \citet{Momose:2016cu} adopted the single-component exponential model for scale-length fit, and the values of \citet{Wisotzki:2016hw} are from two-component modeling.
}
\end{figure}

\begin{figure*}
\includegraphics[width=0.84\paperwidth]{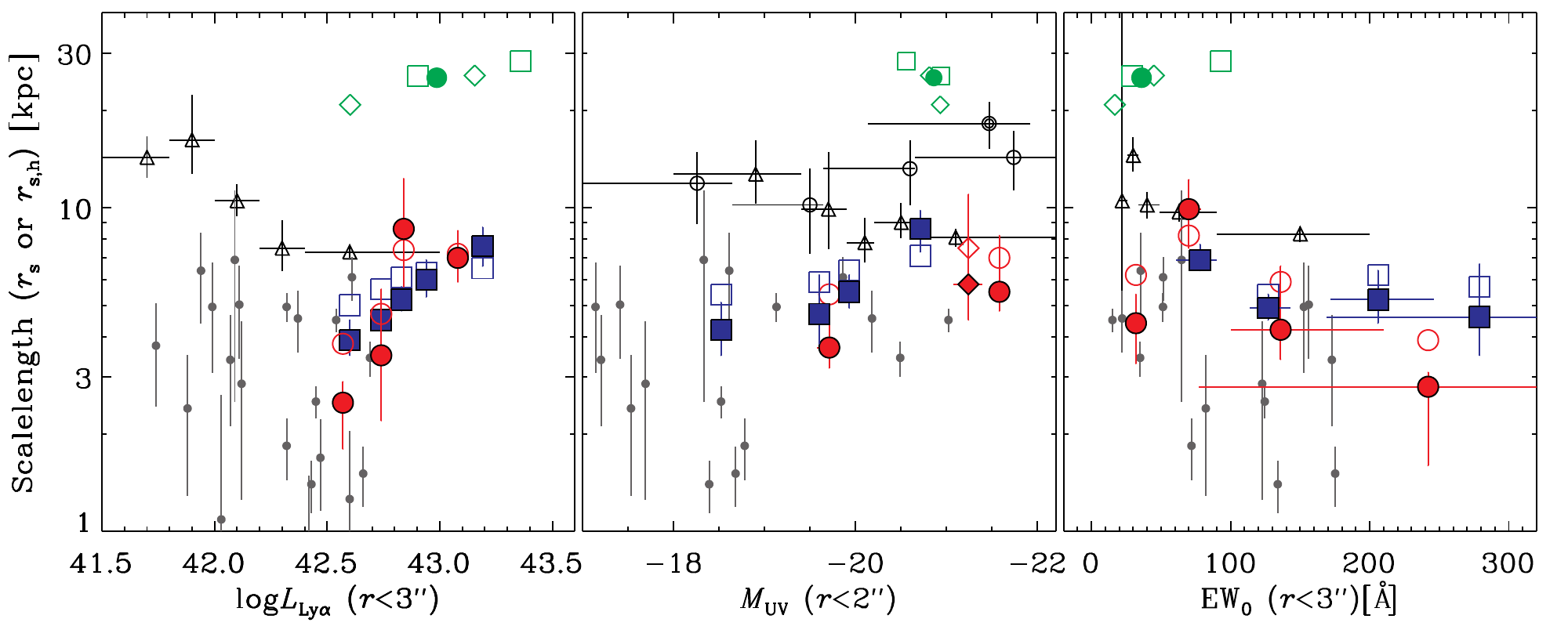}
\caption{\label{fig:slen_prop}
LAH scale-lengths as a function of $\lyalpha$ luminosities (left), absolute UV magnitudes (middle), and Ly$\alpha$ equivalent widths (right). Scale-lengths are measured by fitting the azimuthally averaged surface brightness profile to a single-component exponential model and a two-component model, denoted as $r_{\rm s}$ (open symbols) and $r_{\rm s,h}$ (filled symbols), respectively.  In each panel, our PCF and LAB measurements are shown in red and blue, respectively. Also shown are exponential scale-lengths $r_s$ measurements from S11 (green symbols), \citet{Matsuda:2012fp} (black open circles), \citet{Momose:2016cu} (black open triangles), and \citet[dark gray filled circles]{Wisotzki:2016hw}.
We note that we adopt the same aperture ($r=3\arcsec$) as \citet{{Wisotzki:2016hw}} for measuring \lyalpha-related galaxy physical properties. The other studies adopted slightly different apertures (NB-based isophotal apertures for S11 and \citet{Matsuda:2012fp}, and $r=1\arcsec$ apertures for \citet{Momose:2016cu}).
}
\end{figure*}

\subsection{Local Environments}

In Figure~\ref{fig:slen_env}, we show the scale-lengths of the LAE subsamples as a function of local overdensity, in comparison to the exponential scale-lengths measured by \citet{Matsuda:2012fp} and by \citet{Momose:2016cu}.  
The exponential and two-component scale-lengths are indicated as open and filled color symbols, respectively. 
As detailed in Section\,\ref{sec:measure_overdensity}, we adjust all measured overdensities in this work and in \citet{Matsuda:2012fp} to the benchmark $\delta^{\rm 15\,Mpc}_{\rm sim}$, i.e., the expected three-dimensional overdensity measured in a top hat (15\,Mpc)$^3$ volume.
The similar ranges of the adjusted overdensity parameters suggest that our samples contain overdensity regions comparable to those examined in \citet{Matsuda:2012fp}.

We do not find any correlation between measured scale-lengths and overdensity parameters in either of our data sets. 
Regardless of the measured overdensities, all of our subsamples have LAH sizes of $\sim4$\,--\,$7$\,kpc,  comparable to the results of the field LAEs ($\delta\approx 0$) reported by \citet{Momose:2016cu}. 
The lack of correlation with overdensity parameter is inconsistent with the trend found in \citet{Matsuda:2012fp}.
Although our results are based on different galaxy samples, the LAE selection technique in \citet{Matsuda:2012fp} is similar to ours, and so are the median properties of their subsamples: i.e., Ly$\alpha$ equivalent widths and continuum and Ly$\alpha$ luminosities.

One notable difference does exist between the samples of this study and that of \citet{Matsuda:2012fp}. 
While each of our galaxy samples resides in a contiguous field and is analyzed separately, the LAE collections in \citet{Matsuda:2012fp} are from four independent fields (GOODS-N, SDF, SXDF, and SSA22) and examined together in stacking.
Although it is unclear how many LAEs in their highest overdensity subsample reside in each field (not published), it is possible that their sample is dominated by one particular field---SSA22 \citep{Yamada:2012bm}.
In SSA22, very large LAHs around UV-luminous galaxies and numerous LABs are found around a significant galaxy overdensity region \citep{Steidel:2000bw,Matsuda:2004iz,Steidel:2010go,Matsuda:2011gb}.
In contrast, while the PCF field contains an overdensity as significant as that in SSA22, no LAB has been identified in the entire field.
The LAB field contains the very luminous LAB (LABd05) identified in \citet{Dey:2005dl} and another confirmed candidate detected in \citet{Prescott:2012ed,Prescott:2013iu}, but they are excluded from the stacking analysis.
It is unclear how the presence of LABs impacted the average LAH characteristics reported by \citet{Matsuda:2012fp}.
If we include two known LABs in our $z\approx2.66$ sample, the LAH scale-length of the high-overdensity subsample increases from $r_{\rm s,h}=5.1^{+1.7}_{-1.2}$ to $6.8^{+1.8}_{-1.3}$, leading to a weak positive correlation between the LAH size and overdensity parameter.
While previous studies suggest that LABs are preferentially discovered near LAE overdensities \citep[e.g.][]{Prescott:2008jg,Yang:2009um,Yang:2010ht,Matsuda:2011gb},
careful and uniform analyses on larger samples of protocluster systems are required to understand the effect of LABs on the derived galaxy Ly$\alpha$ halo properties from a stacking analysis, and to decide if they are outliers or a continuous extension of the LAE population.

\subsection{UV Luminosities}

The average LAH sizes increase with UV luminosity in both of our galaxy samples (Figure~\ref{fig:slen_prop}, middle). 
At the lowest luminosities, $M_{\rm UV}\approx -(18$\,--\,$19)$, the measured scale-lengths range from $4$ to $5$\,kpc, similar to those reported by \citet{Wisotzki:2016hw} for individual LAEs of comparable continuum luminosities. 
However, the LAH size in $r_{\rm s, h}$ doubles from 4.2 to 8.6\,kpc for the lowest to highest luminosity bin in the LAB sample.

To evaluate the significance of this correlation, we use the {\sc safe\_correlate} routine in the \idl\ Astronomy Users Library.
The routine can simulate new realizations of scale-lengths based on the measured values and their uncertainties, and then determine the overall probability that the apparent $r_{\rm s, h}$--$M_{\rm UV}$ trend could be due to statistical fluctuations.
For the LAB sample, the test suggests a low probability of the null hypothesis ($\sim$9\%). 
For the PCF sample, {\sc safe\_correlate} (which is based on the Spearman rank correlation) cannot provide useful insight on the correlation significance because we only have only two subsamples due to the limited dynamics range in the sample UV luminosity.
However, the positive correlation trend between $r_{\rm s, h}$ and $M_{\rm UV}$ still holds.

We also stack 21 LBGs in the PCF field with spectroscopic redshifts -- many of which are also UV-bright LAEs -- and find two-component and exponential scale-lengths of 7.5\,kpc and 5.8\,kpc, respectively.
However, the measurements for this PCF subsample are based on a small number of galaxies with faint \lyalpha\ emission, making the results very uncertain. 
We carry out Monte Carlo realizations by randomly stacking subsets of the LBGs, and find a large spread in the measured correlation lengths. Some realizations yield scale-lengths as large as 12\,kpc, much larger than our formal best fit of 7.5\,kpc or 5.8\,kpc.
A larger LBG sample is crucial to further test the possibility of larger LAHs in LBG populations.
 
At the highest luminosities probed by both samples ($M_{\rm{UV}}\approx-21$), our measured scale-lengths are consistent with those reported by \citet{Momose:2016cu} (triangles in Figure\,\ref{fig:slen_prop}); however, they found larger scale-lengths toward lower luminosities of $M_{\rm{UV}}\approx -(19$\,--\,$20)$. 
We note that \citet{Momose:2016cu} used galaxy properties derived from an $r=1\arcsec$ aperture to subdivide samples.
The same photometry is adopted for the presented $M_{\rm UV}$ values of their subsamples here, which may slightly underestimate the galaxy total UV luminosity.
In Figure~\ref{fig:slen_prop}, we show the single-component exponential scale-lengths measured by S11 as green data points. Different symbols represent their subsamples binned by the level of \lyalpha\ emission or the estimated rest-frame EWs. 
All of their scale-length values are substantially higher.

\subsection{Ly$\alpha$ Luminosities}

We find a positive correlation between LAH size and Ly$\alpha$ luminosity, as is evident in the left panel of Figure\,\ref{fig:slen_prop}.
Hence, it appears that the LAH size depends sensitively on Ly$\alpha$ luminosities; nearly doubling the characteristic size from the lowest to highest luminosity bins.
A test using {\sc safe\_correlate} gives the null hypothesis probability of $21\%$ and $8\%$ for the PCF and LAB samples, respectively.
At the lowest Ly$\alpha$ luminosities ($\log L_{\rm{Ly}\alpha}\approx 42.5$), our measurements are once again fully consistent with the individual size measurements of \citet{Wisotzki:2016hw}. 
In contrast, \citet{Momose:2016cu} reported an opposite trend: LAH sizes decrease with increasing Ly$\alpha$ luminosities; however, the trend becomes pronounced only at low luminosities not probed by our samples. 
Most subsamples from S11 probe the high-luminosity regime near $L_{\rm{Ly}\alpha}\approx10^{43}~{\rm erg~s}^{-1}$. 
Their results show a slight increase of the single-component exponential scale-length with Ly$\alpha$ luminosity. However, their measured scale-lengths are systematically larger than those in any other studies, which may reflect the fact that their galaxies are selected as LBGs and the majority of them are considerably more luminous in continuum emission. 

It is evident in Figure~\ref{fig:slen_prop} (left) that the positive correlation between LAH size and Ly$\alpha$ luminosity is stronger if the two-component scale-lengths $r_{\rm s,h}$ are considered instead of the exponential ones $r_{\rm s}$. 
We investigate the difference of $r_{\rm s,h}$ and $r_{\rm s}$, and find that the presence of strong compact Ly$\alpha$ emission in the central region (presumably originating from the galaxy itself) and large-scale PSFs may lead to either overestimation or underestimation in the single-component scale-lengths.
On the other hand, two-component scale-lengths from our analyses should be largely unaffected.
We refer the readers interested in further details of this argument to Appendix,\ref{sec:psf_effect}.

We suspect that the LAH size dependence on Ly$\alpha$ luminosity and on UV luminosity are likely connected. Because the LAE color selection criteria sets the minimum Ly$\alpha$ EW for any galaxy to be selected as an LAE, more luminous galaxies in continuum need to have higher Ly$\alpha$ luminosities to meet the criteria, thus creating an artificial positive correlation between the UV and Ly$\alpha$ luminosities \citep[see, e.g.][]{Ciardullo:2012gt}.
Indeed, the Ly$\alpha$ luminosities and UV continuum luminosities of individual galaxies in our samples are positively correlated, albeit with large scatter (see Figure~\ref{fig:pcflab_propc}), as can be also seen in the median properties for different subsamples presented in Tables~\ref{pcf_table} and \ref{lab_table}. 
These considerations suggest that the observed correlation of LAH sizes with both Ly$\alpha$ and UV luminosities is at least in part influenced by this selection effect. 

\subsection{Ly$\alpha$ Equivalent Widths}

No clear correlation between LAH sizes and Ly$\alpha$ EWs is found in our sample (Figure~\ref{fig:slen_prop}, right). The null hypothesis probability reaches above $40\%$ for both PCF and LAB samples.
Once again, our measured scale-lengths are similar to those of \citet{Wisotzki:2016hw}, while generally lying below the values reported by \citet{Matsuda:2012fp} and \citet{Momose:2016cu}. 
The former observed a weak trend of lower-EW galaxies having slightly larger halos (8.3 and 10.5\,kpc for the highest- and lowest-EW sample, respectively).
The measurements from S11 populate the low end of the EW range because they are UV-luminous LBGs, and their large scale-lengths likely reflect their high continuum luminosities.

Based on the above considerations, we postulate that UV-bright galaxies are likely to have larger LAHs, while Ly$\alpha$ luminosities and EWs, which themselves correlate with UV luminosities, produce similar or weaker correlations with the scale-lengths.
This interpretation is similar to that suggested by \citet{Feldmeier:2013fx} and \citet{Momose:2016cu}, that the galaxies whose properties are closest to those of LBGs appear to have the most extended Ly$\alpha$ halos. 
LBGs, compared to LAEs, generally have higher UV continuum luminosities, EWs, and redder spectral slopes.
In this scenario, the general discordance between different measurements can be partly explained: the \citet{Wisotzki:2016hw} measures lie lower than most because their galaxies are the least UV luminous; the very large scale-lengths measured by S11 reflect the fact that their galaxies are much more continuum luminous than other samples. 

We note that the EWs in our samples, measured individually and on average, are generally within the limit of the ``Case B'' recombination dust-free assumption, i.e., $\lesssim$240\,\AA\ \citep{Charlot:1993cb}.
Although the subsamples with the strongest NB$-$BB color apparently show photometric EWs exceeding $\sim$200\,\AA, their uncertainties are significantly larger due to poor constraints on the UV continuum near $\lambda_{\rm rest}\approx1220$\,\AA. Although our spectroscopic observations do not indicate AGN contamination  in our samples, it is possible that some galaxies in our samples may contain AGNs. The median-combination stacks we use should reject such sources as long as the majority of LAEs do not harbor AGNs. 
Previous studies also suggest a relatively low AGN contribution to the LAE populatio at $z<4$ \citep[e.g.][]{Gawiser:2006fv,Nilsson:2009ib}. Therefore, there is no clear evidence of other \lyalpha\ generation mechanisms rather than the recombination radiation from star formation.

\section{The Origin of $\lyalpha$ Halos: Discussions and Comparison with Theoretical Models}\label{sec:discussion}

\begin{figure*}[!htbp]
\includegraphics[width=0.80\paperwidth]{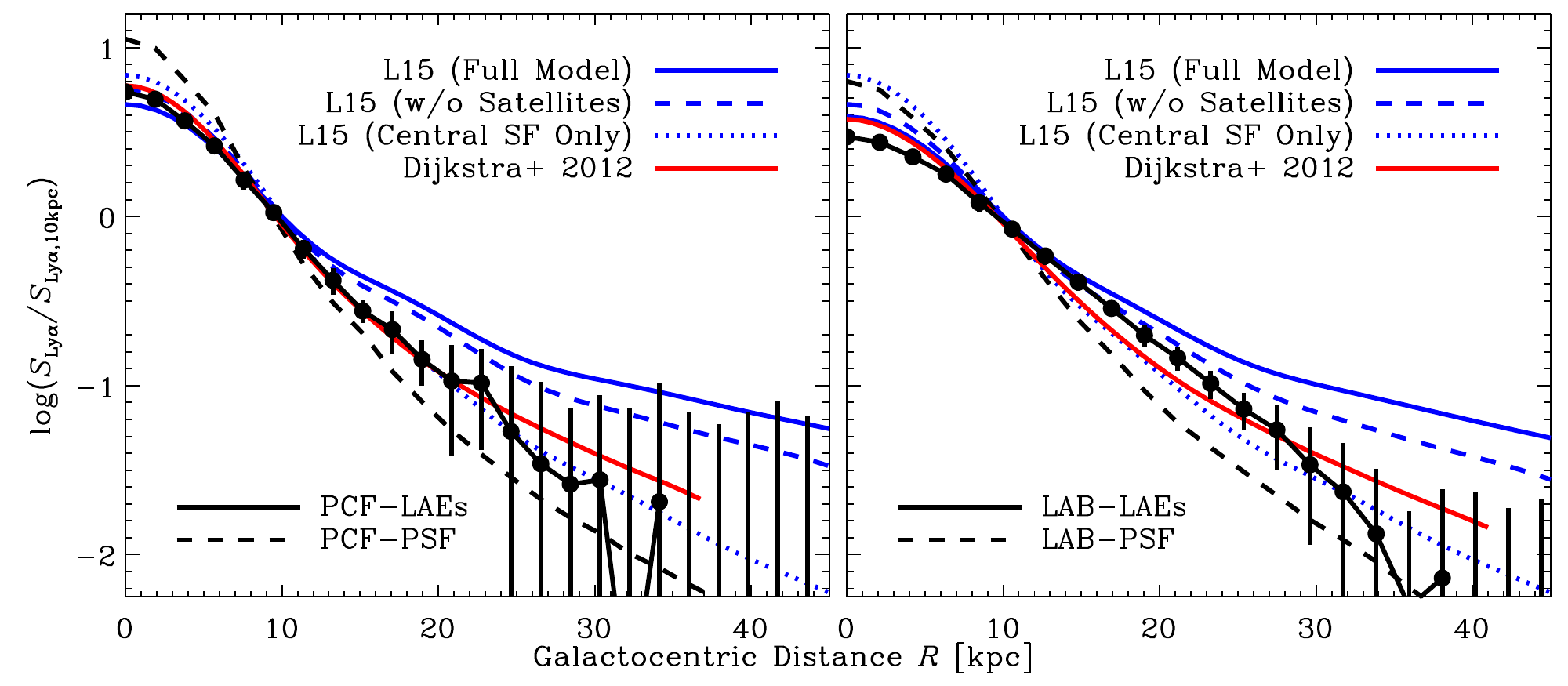}
\caption{\label{fig:compmodels}
Measured LAH radial profiles are compared with theoretical predictions.  All curves are normalized at 10~kpc to compare the slope at larger radii. Our measurements, shown in black, are based on the full LAE samples in the PCF (left) and LAB (right) field. The PSF measured from the image is also indicated in gray. Theoretical models are convolved with the PSFs then normalized at 10\,kpc for fair comparison. 
Three models presented in \citet{Lake:2015gm} are indicated in blue; their full model (solid line) includes star formation in central and satellite galaxies, diffuse star formation in extended regions, and gravitational cooling. Their model without satellite galaxies is shown as a dashed line, while their Ly$\alpha$ profiles arising from the central galaxy are marked as dotted lines. The model prediction given in \citet{Dijkstra:2012ju} is shown in red. In both samples, the measurements closely match the theoretical expectations for Ly$\alpha$ photons arising from central star formation, leaving only little room for possible contributions from other potential sources (see the text for further discussion).
}
\end{figure*}

Apart from the fact that the resonant scattering nature of \lyalpha\ photons provides a mechanism to produce extended LAHs, there is no general consensus in the literature on the expected LAH characteristics.. 
While there is no doubt that at least some \lyalpha\ emission originates from the central star formation of the galaxy itself, \lyalpha\ photons emerging from the outer halo may still be contributed by multiple other sources: namely, star formation in satellite galaxies and diffuse gas around them, and gravitational cooling radiation from infalling gas.
Their relative importance should sensitively depend on the total halo mass, which sets both the number of satellite galaxies and the rate of gas infall responsible for gravitational cooling radiation \citep{Rosdahl:2012bt}.
However, different studies predict a wide range of $\lyalpha$ halo luminosities resulting from gravitational cooling, even at a fixed halo mass \citep[e.g.][]{FaucherGiguere:2010dd}. 
This results from the diverse physical conditions in the gas and complexities of modeling the radiative transfer.
While the strength and spatial distribution of different ionizing sources are critical to determine LAH properties, their effectiveness are also related to the distribution and kinematics of gas and dust -- which scatters and absorbs Ly$\alpha$ photons, respectively -- in the ISM and CGM \citep[hereafter D12]{Dijkstra:2012ju}. 
Viewing angles may also play an important in the LAH appearance \citep{Laursen:2007kl,Verhamme:2012kb}.

In this work, we compare our measurements with two recent models from D12 and \citet[hereafter L15]{Lake:2015gm}. Together, these models bracket the two opposite ends of the possibilities. 

The first model, proposed by D12, assumes that the central star formation is the sole power source of LAHs. 
Ly$\alpha$ photons are produced in the galaxy then propagate through an outflowing, cold, clumpy CGM.
The distribution and kinematics (i.e., outflow velocity field, H\,{\sc i} column density, and degree of clumpiness as a function of galactocentric radius) of the CGM is set to reproduce the $z\approx3$ halo structure derived using observations of quasi-stellar object absorption lines as a function of impact parameter \citep{Steidel:2010go}.

Based on Monte Carlo simulations of radiative transfer for Ly$\alpha$ photons scattering through the CGM, D12 argued that significantly extended LAHs may be produced if the velocity field of gas clouds reaches a maximum at $r\sim 10$~kpc, beyond which clumps decelerate. Their best model produces a clearly extended LAH, but  not as extended as that observed by S11. 

In contrast to the D12 model, L15 takes into account all potential sources of Ly$\alpha$ production, including star formation from central and satellite galaxies, Ly$\alpha$ photons associated with diffuse UV background surrounding these galaxies, and gravitational cooling radiation. 
L15 simulated nine LAEs at $z=3.1$ using an adaptive mesh refinement hydrodynamical simulation of galaxy formation within a cosmological volume. This work re-simulates galaxies at higher resolution, reaching scales of $\approx$120\,pc and mass resolution of $2\times10^7\,M_\odot$. 
They model the \lyalpha\ emission from these halos using the Monte Carlo radiative transfer code of \citet{Zheng:2002jr}, and create an average \lyalpha\ surface brightness profile by combining the LAH models of individual galaxies. 
Each of the emergent Ly$\alpha$ photons is tagged according to its origin, which enables the authors to examine the relative importance of different components at different radii. 
L15 concluded that a significant fraction of their simulated LAHs can come from off-center star formation or gravitational cooling -- contributing roughly equally -- as their Ly$\alpha$ profile from the central SF falls steeply. 
The surface brightness profile from L15 gives an impressively good fit to the \citet{Momose:2014fe} measurements except that it slightly overpredicts at $r>10$~kpc. They further noted that the off-center SF may be suppressed because substantial SF in off-center locations would also lead to an extended UV continuum emission, which is not seen in the \citet{Momose:2014fe} data. Exclusion of off-center SF brings their model to an even better agreement with the observational measurement. 

The caveat of the D12 model is that it only considers photons arising from the central SF propagating through an outflowing medium, and ignores the possibility of an off-center SF and inflow of cold gas. 
On the other hand, L15 used simulated galaxies from hydrodynamic galaxy formation simulations. 
The galaxy formation simulations include feedback from supernova explosion, which drives outflow, and the Ly$\alpha$ radiative transfer modeling is entirely based on the gas distribution produced by the simulations. The model has both outflowing and infalling components, which is supposed to be a more realistic presentation of  anisotropic gas density and velocity distributions.
The presence of ubiquitous outflows in high-$z$ galaxies has been firmly established by observations \citep[e.g.][]{Steidel:2010go} and, as shown by D12, significantly impacts 
the emergent \lyalpha\ profiles.
However, detailed discussions of the difference in the ISM/CGM distributions at radii relevant for the LAH phenomenon are beyond the scope of this paper.

Keeping in mind the differences between these models, we proceed to compare them with the measured LAH surface brightness profile.
In Figure~\ref{fig:compmodels}, we show the radial profile measurements for the full LAE sample (black) and that for point-like sources (gray) in the PCF and LAB field.  
The full L15 model, including central and off-center SF and gravitational cooling, is illustrated by the solid blue line. 
Two variants of their model, by excluding the off-center SF or including only the central SF, are presented as dashed or dotted lines, respectively.
The D12 model is shown by a red solid line. 
Because we are mainly interested in comparing the large-scale behavior and not in evaluating the mean EWs measured (assumed) in different samples (models), all curves are normalized at $r=10$\,kpc. 
We also have convolved the model predictions to match the PSF broadening present in the data. 

The D12 model predicts a remarkably good fit to the PCF measurement at $r=0$\,--\,$30$\,kpc, but slightly underpredicts the LAB profile, and declines with a shallower slope.
The full model of L15 (solid blue line) overpredicts the surface brightness at $r>10$~kpc and also falls more gradually than the data. 
Within the range of $r=10$\,--\,$30$\,kpc, our PCF sample is best described by their ``central SF only'' model while their model that includes gravitational cooling but without off-center SF (dashed line: labeled `without satellites' in Figure~\ref{fig:compmodels}) also lies well above our measurement. 
The LAB profile lies somewhere between their ``central SF only'' model and ``without satellites'' model; however, the slope is more similar to the former. 
Our results indicate that the measured profiles are very close to the models expected from Ly$\alpha$ photons arising from central SF then scattered out to large radii, leaving only a little room for possible contributions from off-center SF (satellites and background diffuse SF) and gravitational cooling. 

At large galactocentric distances ($r>10$\,kpc),  other mechanisms such as SF in satellites and gravitational cooling may contribute to the LAH phenomenon; however, it is difficult to speculate which process may dominate. 
In principle, a large contribution from satellite SF would result in similarly extended halos in the UV continuum images. 
Such a feature is not observed in our UV images; however, similar to that discussed by \citet{Momose:2016cu}, robust detection of the feature requires both  exceedingly low surface brightness sensitivities and a better control on sky subtraction precision. 
In particular, the latter is difficult to achieve as the sky background estimation inherently includes these low-surface-brightness sources, typically leading to over-subtraction, rendering the stacked image to appear more compact than the intrinsic size. Furthermore, a signal from off-center sources would be easily washed away by the median stacking method we adopted. 
Better constraints may come from ultradeep NB observations, which can resolve the morphologies of individual galaxies into compact or diffuse components. 

One useful clue may come from the observed correlation between scale-lengths and the galaxies' physical properties. One clear correlation that emerged from our analysis is that more extended LAHs are found around more UV-luminous galaxies. The trend may be explained in two different ways: first, the physical conditions (the kinematics and distribution of gas and dust) of more UV-luminous galaxies are more conducive to producing extended LAHs; and, second, more UV-luminous galaxies -- forming stars at higher rates -- tend to be hosted by more massive halos \citep{Giavalisco:2001cj,Ouchi:2003fp,Lee:2006fv,Hildebrandt:2007ba}, which are more likely to have a satellite galaxy \citep{Hamana:2004eu,Lee:2009is}.
In the latter scenario, it follows that galaxies in overdense environments are expected to have more pronounced Ly$\alpha$ halos in the average stack as, given everything else fixed, the likelihood of having a companion in proximity is enhanced by the factor that correlates with the overdensity parameter. If such a trend exists, it should have been observed in either of our samples as both fields contain highly significant galaxy overdensities. The complete lack of a trend between LAH size and overdensity parameters (as discussed in Section\,\ref{sec:results}) suggests that satellite populations are not the dominant contributor to the LAH emission.

Interestingly, \citet[][]{Duval:2016cx} recently presented a detailed study of Mrk~1486, a local edge-on disk galaxy  with a large LAH. As a source with one of the highest measured Ly$\alpha$ luminosities and EWs in the LARS sample \citet{Hayes:2013jc}, Mrk\,1486 also has physical properties (stellar mass, age, SFR, and dust content) similar to those of typical high-redshift LAEs \citep{Hayes:2014jv}. Based on multiple line diagnostics, they concluded that Ly$\alpha$ photons are likely photoionized inside the disk then scattered in our direction by neutral gas in bipolar outflows. The physical picture presented by  \citet[][]{Duval:2016cx} is qualitatively similar to ours and thus lends further support to our main conclusions. 
\section{Summary}\label{sec:summary}

In this paper, we report the robust detection of diffuse Ly$\alpha$ emission around high-redshift star-forming galaxies. Taking advantage of two large spectroscopic/photometric samples of galaxies at $z\approx2.66$ and $z\approx3.78$, we have examined how the LAH sizes correlate with the physical properties of galaxies, with rigorous tests of possible systematics in constructing stacked Ly$\alpha$ images and measuring LAH characteristics.  Our main results are described as follows.\\

\begin{enumerate}

\item In our full samples and most of our subsamples, the stacked \lyalpha\ image is significantly more extended than the UV continuum image of the same galaxies, unambiguously confirming the presence of diffuse Ly$\alpha$ emission. Typical sizes of LAHs in our samples are relatively modest at 4\,--\,8~kpc, in good agreement with recent measurements of individual galaxies \citep{Wisotzki:2016hw}. Very large LAHs ($>15$\,kpc) similar to those reported by \citet{Steidel:2011jk} are not detected.

\item We examine how LAH sizes---measured as an exponential scale-length---depend on galaxy properties: namely, UV luminosity, $\lyalpha$ luminosity, and EWs, and the local environment. In contrast to \citet{Matsuda:2012fp}, we find no correlation between LAH size and local environment -- measured in galaxy overdensity -- even though both of our samples contain significantly overdense structures, which will likely evolve to massive galaxy clusters by the present epoch.
The reason for the discrepancy remains unclear.
The strongest correlation is found with UV luminosity in our data. The same trend becomes even stronger when combined with other measurements in the literature. 
We conclude that the physical processes that determine LAH appearance likely correlate strongly with a galaxy's UV luminosity. The observed LAH trends with other galaxy parameters may be driven by the fact that they themselves weakly correlate with UV luminosity.

\item We compare the LAH profiles measured in our data with recent theoretical predictions.  A simple model in which Ly$\alpha$ photons originate from  central star formation, then resonantly propagate outward in a clumpy outflowing medium, is in a reasonable agreement with our measurements. 
The implication is that other potential producers of Ly$\alpha$ photons, such as low-level star formation occurring in off-center locations and radiative cooling of collisionally heated H\,{\sc i} gas (gravitational cooling), may be at best minor contributors to the extended Ly$\alpha$ emission around normal star-forming galaxies.

\end{enumerate}

\acknowledgments

We thank the referee for a careful reading of the manuscript and for suggestions that helped improve this paper. 
This paper presents data obtained at the W. M. Keck Observatory (NASA proposal ID numbers 2014A-N116D and 2015A-N142D) and the Mayall 4m telescope of the Kitt Peak National Observatory (NOAO proposal ID numbers 2012A-0454, 2014A-0164, and 2014B-0626). 
We are grateful to the NASA Keck and NOAO Time Allocation Committees for granting us telescope time and to the staff of the W. M. Keck Observatory and the Kitt Peak National Observatory. 
Based in part on data collected at Subaru Telescope, which is operated by the National Astronomical Observatory of Japan (Subaru proposal ID \# o07123, o08226).
Part of the observations reported here were obtained at the MMT Observatory, a joint facility of the University of Arizona and the Smithsonian Institution. 
We thank Y.-K. Chiang et al. for sharing the results of their protocluster studies to facilitate our overdensity calibration.
We also thank Mark Dijkstra for providing insightful comments. 
A.D.'s research is supported by the National Optical Astronomy Observatory (NOAO). NOAO is operated by the Association of Universities for Research in Astronomy (AURA), Inc. under a cooperative agreement with the National Science Foundation.

\appendix

\section{Calculations of Ly$\alpha$ Luminosities and Equivalent Widths}
\label{sec:flya}

Here, we briefly describe the procedure adopted to compute the rest-frame \lyalpha\ equivalent widths and line luminosities. 
Assuming a galaxy at redshift $z$ of which the continuum slope is described by a power law of $f_{\lambda}\propto\lambda^\beta$, the equivalent monochromatic flux density for a given passband containing both UV continuum and Ly$\alpha$ emission line can be expressed as follows,
\begin{equation}
f_{\rm AB} \equiv 10^{-0.4(m_{\rm AB}+48.6)} = \frac{F_{\rm{Ly}\alpha}}{B} + f_{\rm{cont}}Q
\end{equation}
where $F_{\rm{Ly}\alpha}$ represents the integrated Ly$\alpha$ flux in units of erg\,s$^{-1}$\,cm$^{-2}$, and $f_{\rm{cont}}$ is the intrinsic specific continuum flux density near the Ly$\alpha$ line in units of erg\,s$^{-1}$\,cm$^{-2}$\,Hz$^{-1}$.
$B$ is the effective bandwidth measured in units of Hz, defined as
\begin{equation}
B \equiv \frac{c\int \lambda^{-1} \mathcal{R} (\lambda) d\lambda}{\mathcal{R}(\lambda_0)~\lambda_0}.
\end{equation}
Here, $\lambda_0$ is the \lyalpha\ wavelength in the observer's frame and $\mathcal{R}(\lambda)$ is the system total throughput\footnote{The system throughput here is measured in units of photon$^{-1}$, representing the system response function defined by the filter, CCD quantum efficiency, and telescope/atmospheric transmission. 
We assume an airmass of 1.0 and adopt the transmission function of each filter from its respective website:
{\it IA445}, \url{http://www.awa.tohoku.ac.jp/astro/filter.html};
${\it WRC4}/B_W/R/I$, \url{http://www.noao.edu/kpno/mosaic/filters}.}.
The dimensionless $Q$ factor is defined as
\begin{equation}
Q(z,\beta) \equiv 
\frac
{\int e^{-\tau_{\rm IGM}} (\lambda/\lambda_0)^{2+\beta}\lambda^{-1} \mathcal{R}(\lambda) d\lambda}
{\int \lambda^{-1} \mathcal{R}(\lambda) d\lambda},
\end{equation}
where $\tau_{\rm{IGM}}$ denotes the effective optical depth of the IGM as a function of wavelength.
We adopt the IGM transmission prescription from \citet{Inoue:2014cw}, which is based on the updated statistics of the intervening absorption systems, and is comparable to the model of \cite{Meiksin:2006kq} longward of the Lyman continuum break. 
However, it generally predicts higher transmission than the commonly adopted model of \citet{Madau:1995jb}.
The $Q$ factor represents two observational effects: the absorption by the IGM neutral hydrogen averaged over the bandpass, and the conversion from the specific flux density of the continuum near \lyalpha $f_{\rm cont}$ to its AB magnitude.
If the UV continuum slope is $\beta=-2$, the $Q$ factor will be unity for a passband spanning the rest-frame wavelengths $\lambda>1216$\,\AA. 

When we have both narrow- and broadbands (NB/BB) observations covering the \lyalpha\ line and adjacent UV continuum, the integrated line flux $F_{\rm{Ly}\alpha}$ and the continuum specific flux density $f_{\rm{cont}}$ are expressed as
\begin{eqnarray}
F_{\rm Ly\alpha}&=&
\frac{B_{\rm BB}B_{\rm NB}}{Q_{\rm BB}B_{\rm BB}-Q_{\rm NB}B_{\rm NB}}\\
&&\times(Q_{\rm BB}f_{\rm AB,NB}-Q_{\rm NB}f_{\rm AB,BB}) \nonumber \\
f_{\rm cont}&=&\frac{B_{\rm BB}f_{\rm AB,BB}-B_{\rm NB}f_{\rm AB,NB}}{Q_{\rm BB}B_{\rm BB}-Q_{\rm NB}B_{\rm NB}}
\end{eqnarray}
The rest-frame equivalent width is then given as:
\begin{equation}
{\rm EW}_{0}=\frac{{\rm EW_{obs}}}{1+z} =
\frac{\lambda_0^2}{c}\frac{F_{\rm Ly\alpha}}{ (1+z)f_{\rm cont}}
\end{equation}
These quantities derived for the NB/BB datasets in the PCF and LAB fields are listed in Table\,\ref{tab:qbtable}. 
By shifting the redshift of \lyalpha\ across the effective range selected by the NB filter and adjusting $\beta=-(1$--$2)$, we find that the relative error of continuum subtraction is below 10\%.

\begin{deluxetable}{cccccc}\tabletypesize{\footnotesize}
\tablecolumns{8} 
\tablecaption{$Q$ and $B$ Values for Our Data Sets\label{tab:qbtable}}
\tablewidth{0pt} 
\tablehead{
\multicolumn{1}{c}{NB-BB/}&
\multicolumn{1}{c}{$\beta$}&
\multicolumn{1}{c}{$Q_{\rm NB}$}&
\multicolumn{1}{c}{$B_{\rm NB}$}&
\multicolumn{1}{c}{$Q_{\rm BB}$}&
\multicolumn{1}{c}{$B_{\rm BB}$}\\
\multicolumn{1}{c}{Redshift}&
\multicolumn{1}{c}{}&
\multicolumn{1}{c}{}&
\multicolumn{1}{c}{Hz}&
\multicolumn{1}{c}{}&
\multicolumn{1}{c}{Hz}
}
\startdata	
${\it WRC4}-R$ 	&  -1.0 & 0.763 & \nodata  & 1.070 & \nodata\\
$z=3.790$ 		&  -1.5 & 0.763 & $3.726\times10^{12}$ & 1.006& $1.284\times10^{14}$\\
			&  -2.0 & 0.762 & \nodata & 0.948& \nodata\\
\midrule
${\it IA445}-B_{W}$ &  -1.0 & 0.808 & \nodata & 0.795 & \nodata\\
$z=2.698$ 		&  -1.5 &0.812 & $2.853\times10^{13}$ &0.820 &$1.520\times10^{14}$ \\
 		&  -2.0 & 0.817 & \nodata   & 0.847& \nodata\\		
\enddata
\end{deluxetable}

\section{Comparison of Stacking Methods}\label{sec:stack_effect}

For constructing the Ly$\alpha$ surface brightness profile, previous studies employed different approaches to image masking, scaling/weighting and stacking.
Most studies used a simple sigma-clipping approach, while we also experiment with pixel-repaired images as described in Section\,\ref{sec:defaultstack} in this work. 
Images can be rescaled/weighted differently for the signal-to-noise optimization of the stacked image and outlier rejection. 
Finally, the image stacking can be performed by taking the pixel-wise median or mean, typically on small-sized cutouts centered on each galaxy. 
If Ly$\alpha$ luminosities and LAH profiles vary significantly within a given sample, the light profile from mean stacks, compared with median stacks, may be biased toward sources with higher surface brightness.
We note that any low-level unmasked contaminants of each cutout can artificially enhance the signal, particularly for mean-combined stacks.

For the present work, we employ median image combination of pixel-repaired cutouts without further scaling and weighting. In comparison, S11 used a straight mean image combination with masking, while \citet{Matsuda:2012fp} adopted a median image combination without masking. 
\citet{Momose:2014fe} used two methods: one uses a weighted mean image combination on sigma-clipped galaxy cutouts; the other uses median combination on scaled images.  \citet{Feldmeier:2013fx} first rescaled the NB images to a common flux level, then used a sigma-clipped mean combination with weights proportional to the total NB flux of each galaxy.  

\begin{figure*}[!htbp]
\begin{minipage}{0.68\linewidth}
\includegraphics[width=0.9\linewidth]{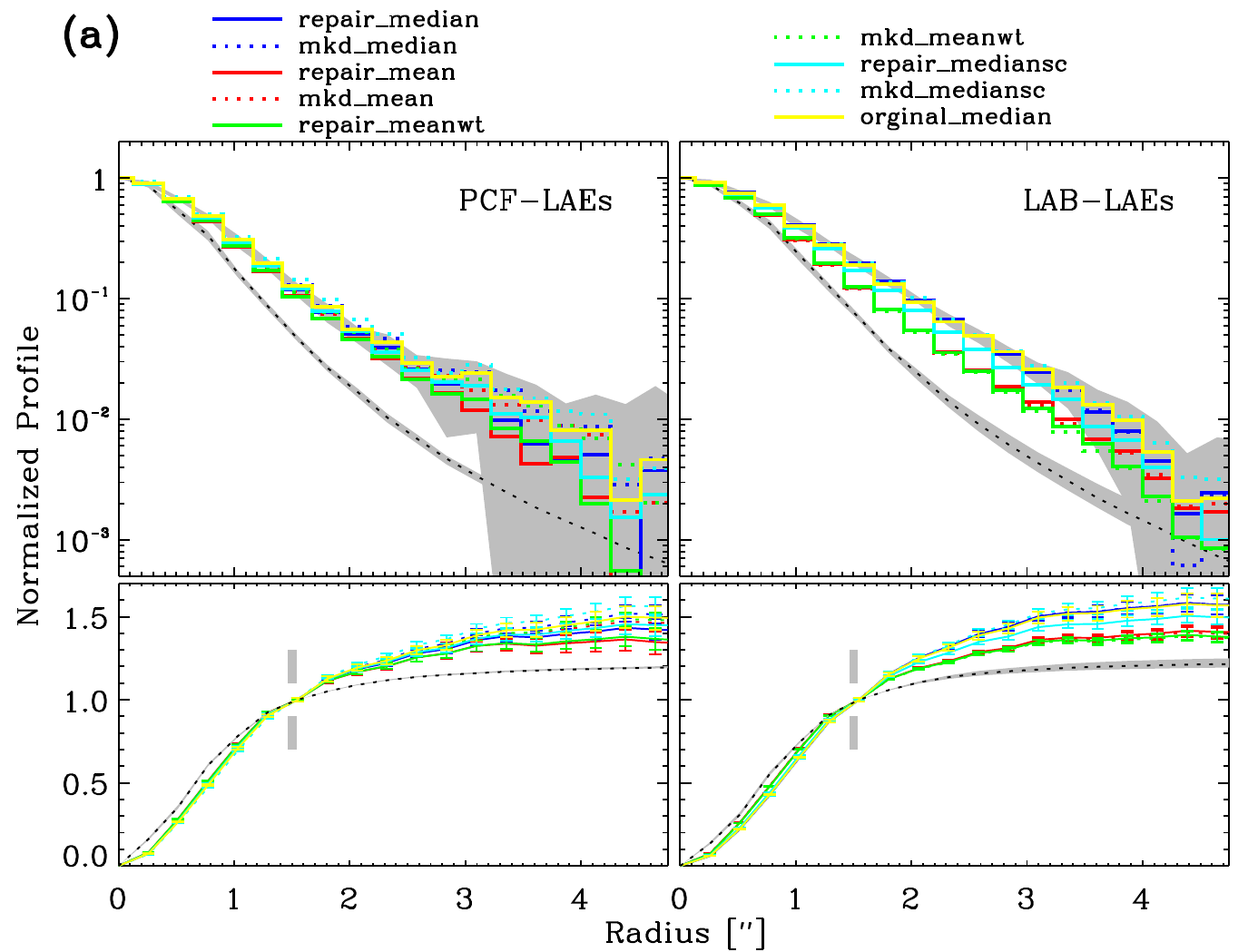}
\end{minipage}
\begin{minipage}{0.32\linewidth}
\includegraphics[width=0.9\linewidth]{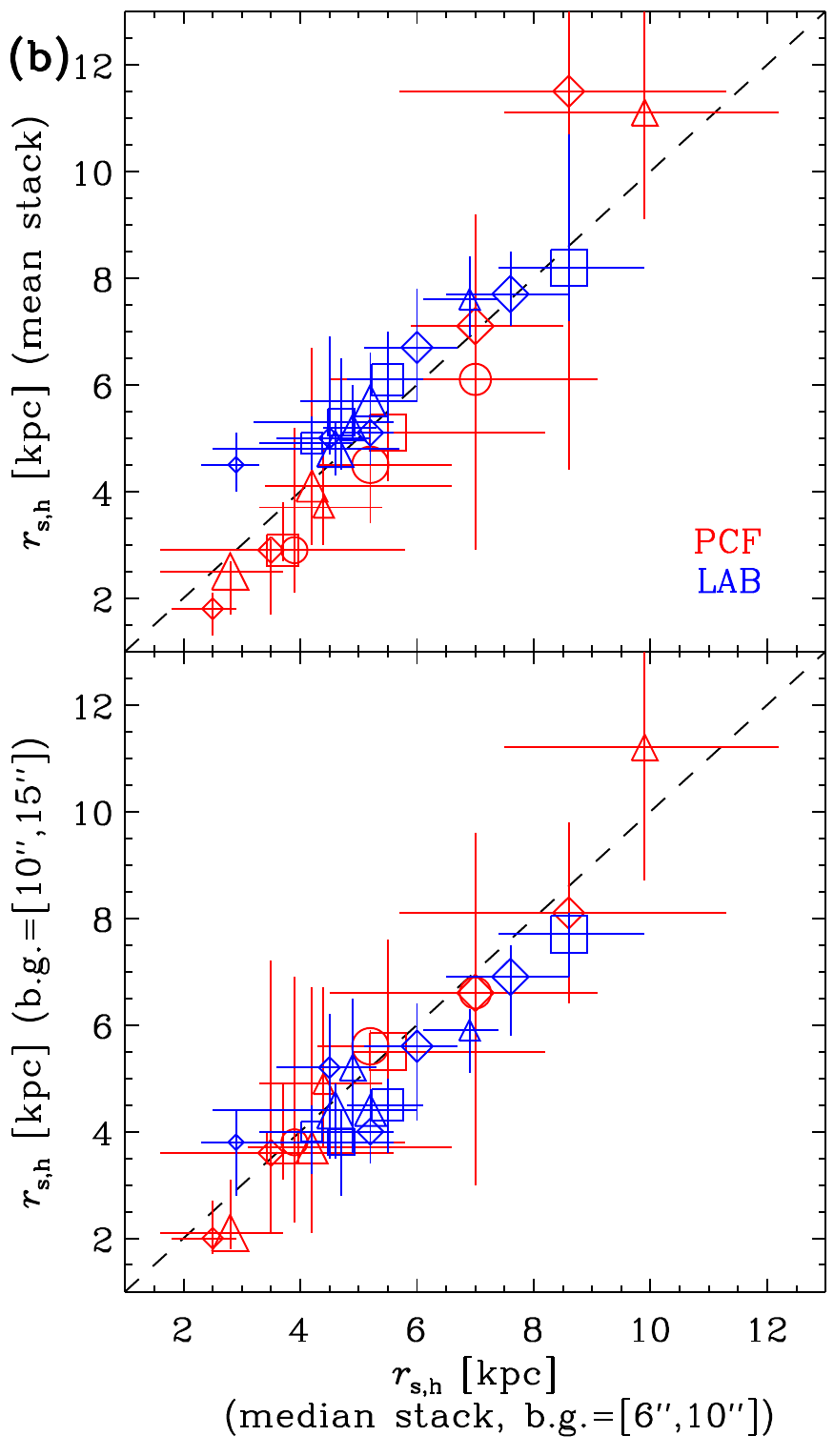}
\end{minipage}
\caption{\label{fig:stackcomplya} \label{fig:bgcomplya} 
(a) \lyalpha\ radial profiles resulted from nine different stacking methods are compared. The full LAE sample is used for stacking in all methods. All differential profiles (top), including the PSF (dotted line), are normalized at the center ($\theta=0$\arcsec), while all cumulative profiles  (bottom) are normalized at $\theta=$5\arcsec. Estimated uncertainties of our fiducial model (repair\_median stacking: blue solid line) are indicated as gray shades. We find that different methods generally return similar slopes in the radial profiles, but with a range of amplitudes. Scale-lengths should provide a reliable metric for the characterization of an average LAH. More discussion about the differences among different methods is given in the text.
(b) LAH scale-lengths $r_{\rm s,h}$ measurements using different image stacking and background estimation methods are shown for the PCF (red) and LAB (blue) subsamples.
In both panels, these samples are defined according to their UV luminosities (square), \lyalpha\ luminosities (diamond), \lyalpha\ equivalent widths (triangle), and LAE overdensities (circle) as defined in Table~\ref{pcf_table} and Table~\ref{lab_table}, respectively. Symbol size increases as the median value of these quantities increases. 
In the top panel, the scale-lengths measured from median- and mean-combined images are compared. 
In the bottom panel, the scale-lengths are measured from median-combined image using different annular regions for sky background estimation. 
In both cases, we find that our scale-length measurements are robust against specific choices we made for deriving average radial profiles.
}
\end{figure*}

In an effort to explore possible systematics inherent to different stacking techniques, we stack our full LAE sample using nine different stacking methods. 
These include 
(i) median combination with masking ({\it mkd\_median}); 
(ii) median combination of repaired images  ({\it repair\_median}): the method adopted in this work; 
(iii) mean combination with masking ({\it mkd\_mean}); 
(iv) mean  combination of repaired images ({\it repair\_mean}); 
(v) weighted mean combination with masking ({\it mkd\_meanwt}); 
(vi) weighted mean combination of repaired images ({\it repair\_meanwt}); 
(vii) median combination of rescaled images with masking ({\it mkd\_mediansc}); 
(viii) median combination of rescaled repaired images ({\it repair\_mediansc}); 
and (ix) straight median combination with no masking ({\it original\_median}). 
For rescaling, we normalize each cutout by the 2\arcsec\ circular aperture flux to the median value of all galaxies in the sample. 

The results of our test are illustrated in Figure~\ref{fig:stackcomplya} where we show the differential (top) and cumulative (bottom) radial profiles constructed using nine methods together with the PSFs in both fields.  All methods clearly detect extended emission compared with the PSF (dashed black line) in both  profiles. The median stacks generally exhibit a slightly broader profile than the mean stacks, which is not unexpected for an intrinsically skewed distribution. However, the radial slope does not change significantly within the range in which the exponential scale-lengths are measured ($r\approx2\arcsec-5\arcsec$), resulting in most models having very similar scale-lengths. More specifically, the  exponential scale-length ranges from $r_{\rm s}=5.4$\,kpc ({repair\_mean}) to 6.9\,kpc ({original\_median}) compared to our fiducial value, $r_{\rm s}=5.8$\,kpc ({repair\_median}) for the PCF sample. 
No significant variation is found in the LAB sample. We also verify that the pixel-repair technique returns a result consistent with the simple masking of pixels. 

Our test suggests that the {\it orginal\_median} method leads to a slightly broader profile. 
The trend is most clearly illustrated from our test on the PCF sample (yellow line in Figure~\ref{fig:stackcomplya}). 
We speculate that the pixel-wise median combination may not be effective enough to remove all contamination from a data set of moderate size, which leads the stacked \lyalpha\ image to contain an enhanced level of flux out to very large radii. 
The scale-lengths, $r_{\rm s,h}$, from the two-component model are also robust against the details of how the image is created. 
In the top panel of Figure\,\ref{fig:bgcomplya}(b), we compare the measured $r_{\rm s,h}$ values of individual subsamples (as discussed in Section\,\ref{sec:results}) using the median and mean stacking methods. 
We find a reasonable agreement within the majority of subsamples. Our results presented in Section\,\ref{sec:results} do not change qualitatively if we adopted  mean stacking instead. 

The fluctuations in amplitude of measured radial profiles  among different stacks are likely also linked to the precision of sky subtraction we are able to perform. As described in Section\,\ref{sec:defaultstack},  a radial profile is measured from a ``contamination-free zero-sky''  image, which is constructed by estimating and subsequently subtracting the sky background value from the stacked image itself.
While this procedure does an excellent job ensuring that the pixels far from the central source are nearly background-free, the background value determined on the stacked image mildly fluctuates depending on the type of cutout images (i.e., {repaired}, {masked}, or {original}) and the choice of image combination ({\it mean} versus {\it median}). 

Small uncertainties in sky background can alter the radial profile appearance, especially in the low-S/N regions. 
However, we find that measured scale-lengths are insensitive to the specific choice adopted for sky background estimation. 
In the bottom panel of Figure~\ref{fig:bgcomplya}(b), we show the $r_{\rm s,h}$ values from our individual subsamples, using two different annular regions $r=$[6\arcsec, 10\arcsec]  and $r=$[10\arcsec, 15\arcsec] to determine background levels. 
The former is adopted as our default method in this work. 
The comparison shows a good consistency in $r_{\rm s,h}$ despite very different choices of the annular regions adopted for sky subtraction.

Based on the comparisons in Figure~\ref{fig:bgcomplya}(b), we conclude that scale-length measurements $r_{\rm s,h}$ offer a robust method to characterize the properties of diffuse \lyalpha\ emission in the observational viewpoint. 
In comparison, the overall surface brightness is sensitive to the manner in which image stacking and sky subtraction are performed, and to the overall shape of the image PSF. We discuss the impact of the latter in Appendix\,\ref{sec:psf_effect}.

\section{Impact of PSF on Measurements of Scale-lengths}\label{sec:psf_effect}

\begin{figure*}[!htbp]
\centering
\begin{minipage}{0.40\linewidth}
\includegraphics[width=1.00\linewidth]{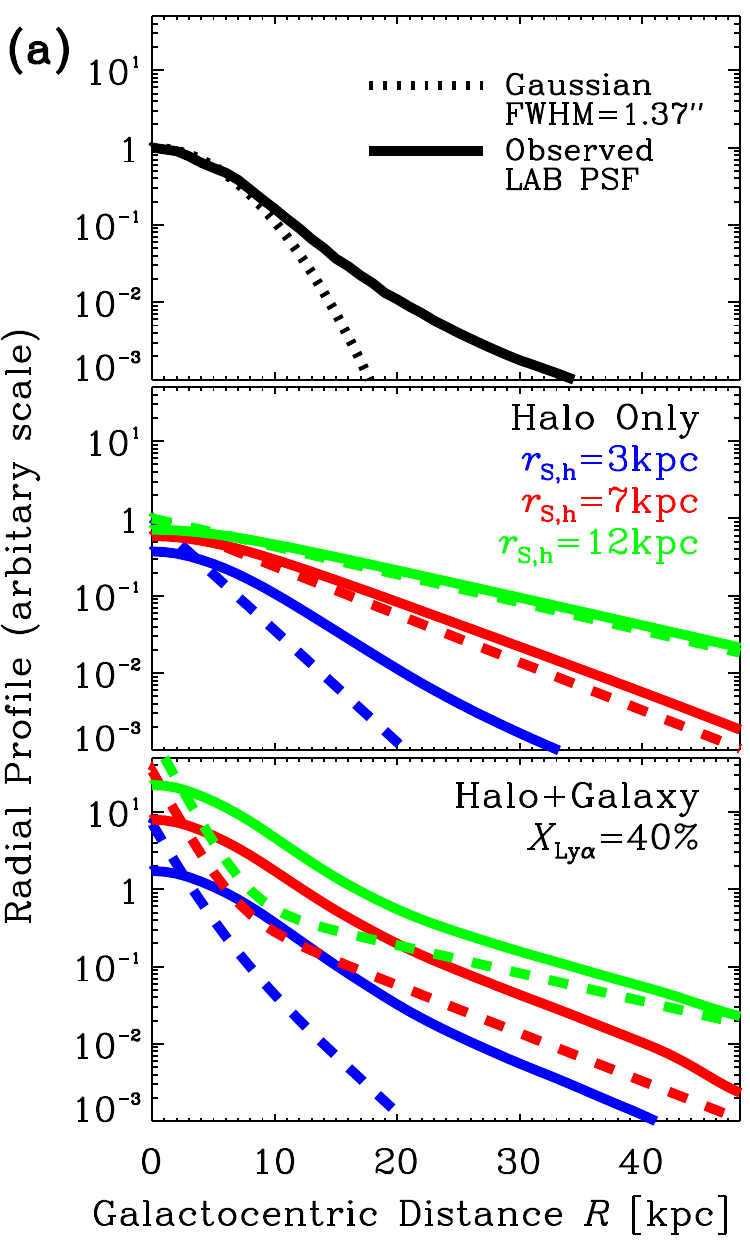}
\end{minipage}
\begin{minipage}{0.50\linewidth}
\includegraphics[width=1.00\linewidth]{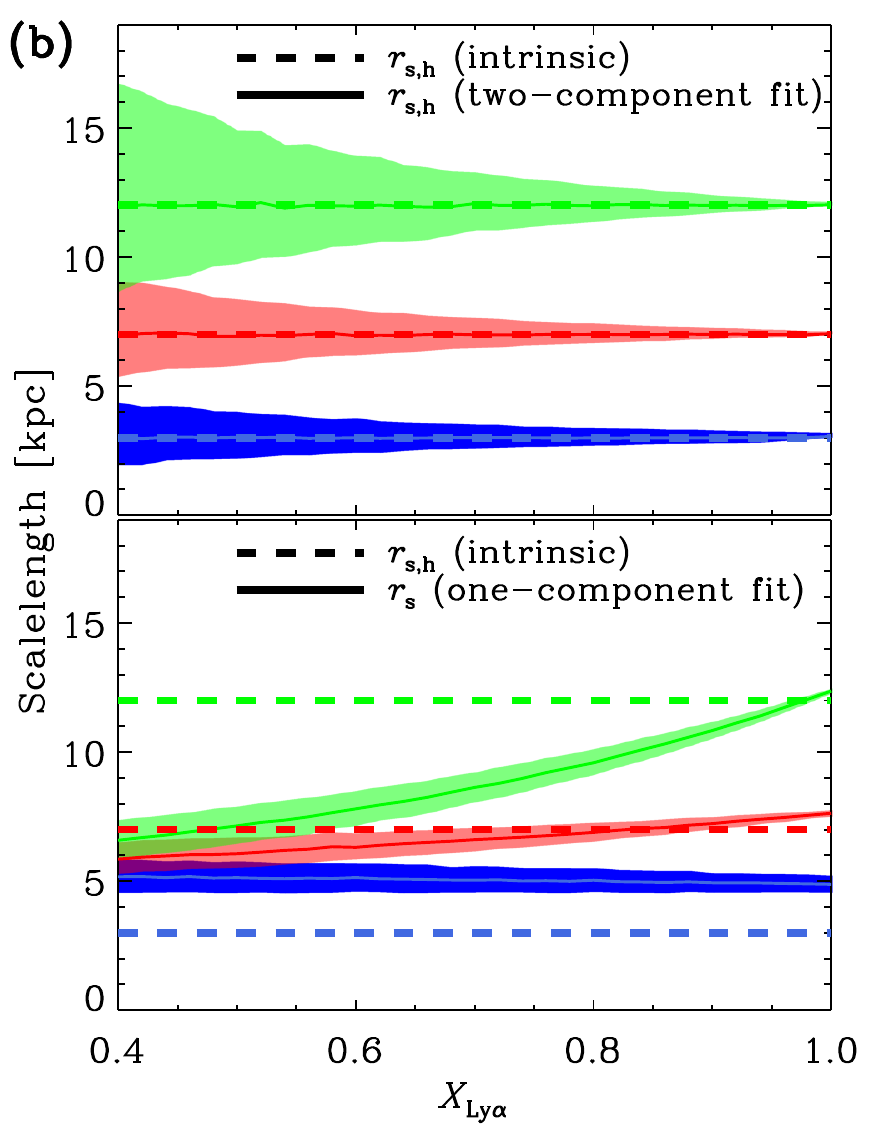}
\end{minipage}
\caption{\label{fig:stackmodelssim}
(a) PSF effects of the Ly$\alpha$ radial profile are illustrated with three LAH toy models.
The Ly$\alpha$ surface brightness is modeled as a superposition of two components, namely, a compact galaxy-like component and a Ly$\alpha$ halo; both are modeled as exponentially declining functions with a scale-length of 1.5\,kpc for the former, and with a scale-length of 3, 7, or 12\,kpc for the latter.
The intrinsic profile is shown as a dashed line, while the ``observed'' profile (solid color line) is created by convolving it with the PSF of our LAB sample.
In the middle and bottom panels, the fraction of Ly$\alpha$ flux originating from the LAH component, denoted as $X_{\rm{Ly}\alpha}$, is assumed to be 100\% and 40\%, respectively.
The PSF measured from our data (top panel, solid black line) exhibits a broad wing, and as a result significantly alters the intrinsic profile by scattering centrally emitted photons to large radii. 
(b) We simulate a series of Ly$\alpha$ radial profiles based on our toy models, by varying $X_{\rm{Ly}\alpha}$ from 40\% to 100\% and adding random noise at a level similar to the real data.
The derived scale-lengths ($r_{\rm s,h}$ and $r_{\rm s}$ in the top and bottom panels) are shown as a function of $X_{\rm{Ly}\alpha}$; thin solid lines mark the median values, while color shades are defined by the lower-upper quartile ranges. Dashed lines represent the intrinsic scale-lengths.
The two-component fitting recovers the intrinsic values more robustly than the single-component fitting method. 
}
\end{figure*}

Large-scale PSFs can significantly alter the observed radial profile of the Ly$\alpha$ emission \citep[e.g.,][]{Feldmeier:2013fx}; as a result an LAH exponential scale-length directly measured from the apparent profiles ($r_{\rm s}$ from Equation\,\ref{eq:expo}) may be susceptible to the PSF effect.
Here, we use three toy models to illustrate its potential bias.

In our toy models, the \lyalpha\ emission from a star-forming galaxy consists of two components as described in Equation\,\ref{eq:twoc}: a halo component following an exponential profile with a scale-length of 3, 7, or 12\,kpc; and, a galaxy component which exponentially declines with a scale-length of $r_{\rm s,c}=1.5$~kpc (a typical size of LAEs at our sample redshift).
We define the halo-to-total Ly$\alpha$ flux fraction as
\begin{equation}
X_{\rm{Ly}\alpha} \equiv F_{\rm{Ly}\alpha,\rm{halo}}/( F_{\rm{Ly}\alpha,\rm{halo}}+F_{\rm{Ly}\alpha,\rm{galaxy}})
\end{equation}
where $F$ denotes the total flux of the halo or galaxy component. 
In each of our models, we vary the halo-to-total Ly$\alpha$ fraction, with a ``halo-only'' model corresponding to $X_{\rm{Ly}\alpha} = 100$\%.
Then we simulate the apparent Ly$\alpha$ radial profiles by convolving intrinsic profiles with the homogenized PSF of our LAB data set at $z=2.66$.
In Figure~\ref{fig:stackmodelssim}(a), we present the adopted PSF in the top panel, along with the simulated intrinsic and apparent \lyalpha\ radial profiles at two halo-to-total Ly$\alpha$ fractions ($X_{\rm{Ly}\alpha}=100\%$ or 40\%) in the middle and bottom panel.
Three toy models are shown in different colors. 

In the ``halo-only'' ($X_{\rm{Ly}\alpha}=100\%$) scenario, the apparent profile of the compact LAH model (3\,kpc) closely follows the PSF.
However, for the 12\,kpc LAH model, the observed profile is very similar to the intrinsic profile except within the central 5\,kpc.
This suggests that, when all of the Ly$\alpha$ emission originates from an LAH, the measured scale-lengths should be relatively insensitive to the PSF as long as it falls off more steeply than the LAH.
On the other hand, if the PSF falls off more shallowly than the halo itself and the PSF is ignored in the analysis, the scale-length will be overestimated because the profile is essentially determined by the PSF.

If the Ly$\alpha$ emission has two components (i.e., a compact emission region associated with the galaxy and a more spatially extended halo), then fitting the profile with a single exponential may result in significant underestimations of the halo size. 
This is illustrated in the bottom panel of Figure~\ref{fig:stackmodelssim}(a), in which 40\% of Ly$\alpha$ emission originates from the galaxy component.
While the compact LAH models (3 or 7\,kpc) resemble the PSF at large radii, the observed profile of our 12\,kpc LAH model steepens significantly from the intrinsic profile between 10\,kpc and 40\,kpc.
In contrast to the ``halo-only'' scenario, this shows that the integrated \lyalpha\ light profile is more subject to the PSF effect when the contribution from a compact Ly$\alpha$ component---resembling the PSF---becomes prominent.

We perform the single-component and two-component exponential fitting to a series of simulated profiles to test their capability to recover the intrinsic halo scale-lengths in the identical manner to the real data.
We generate the simulated Ly$\alpha$ profiles by varying $X_{\rm{Ly}\alpha}$ from 40\% to 100\% --- a range consistent with that inferred from our own data, and with similar measurements on individual LAEs reported by \citet{Wisotzki:2016hw}.
Then, artificial noise is added to the simulated profiles at a level of peak S/N=100 similar to that observed in the real data.

In Figure~\ref{fig:stackmodelssim}(b), we show the distributions of LAH scale-lengths derived from two-component fitting ($r_{\rm s,h}$, top panel) and single-component fitting ($r_{\rm s}$, bottom panel) for the three toy models as a function of halo-to-total Ly$\alpha$ flux fraction $X_{\rm{Ly}\alpha}$.
The median values of $r_{\rm s,h}$ successfully recover the intrinsic scale-lengths in all models, though the scatter becomes larger as $X_{\rm{Ly}\alpha}$ decreases.
The result is not surprising:
accounting for the PSF and the contribution from the compact Ly$\alpha$ component are critical to a robust estimate of the halo extent;
increased uncertainties in derived scale-lengths at lower $X_{\rm{Ly}\alpha}$ are the result of decreased \lyalpha\ flux from the halo component as the added noise is normalized by the S/N of the total flux (i.e., galaxy and LAH). 
On the other hand, the single-component scale-length $r_{\rm s}$ recovers the intrinsic value only if: (i) the halo profile is broader than that of the PSF ($r_{\rm s,h}\gtrsim5$\,kpc); and (ii) the compact \lyalpha\ component is not prominent ($X_{\rm{Ly}\alpha}\gtrsim80\%$).
Even for the largest LAH, measured scale-lengths can depend sensitively on the halo fraction.

The bias of using $r_{\rm s}$ to characterize LAHs would be less significant if the PSF falls off steeply at large radii similar to a Gaussian function. 
In the top left panel of Figure~\ref{fig:stackmodelssim}(a), we show the PSF in the LAB data along with a Gaussian profile of the same FWHM.
Realistically, the PSFs in both of our data sets are approximated by Moffat profiles with $\beta\approx3$\,--\,$4$, which are typical in ground-based imaging \citep[e.g.][]{Dey:2014bv}.
This result not only highlights the importance of quantifying the large-scale PSF in the LAH measurements, but also indicates the vital role of PSF homogenization in stacking-based analyses.
High-precision measurements of large-scale PSFs and subsequent PSF homogenization performed on galaxy cutouts accordingly are key to robustly measuring the LAH sizes.

Without the details of large-scale PSFs, we are unable to quantify how significant the above biases could be in previous studies adopting $r_{\rm s}$ to characterize LAH sizes.
The two-component analyses from both our study and \citet{Wisotzki:2016hw} (see their Figure 14) indicate that Lya flux is dominated by the halo component: i.e., $X_{\rm{Ly}\alpha}<0.5$ is rare even among LAEs. In such a case, a less severe bias in the estimation of $r_{\rm s}$ is expected.
However, the uncertainties in current measurements are still large and prohibit us from offering a conclusive argument.
Nevertheless, our simulation does show that one needs to exercise caution when interpreting the LAH scale-lengths measured from different methods and comparing the values derived from different galaxy samples.

\bibliographystyle{aasjournal}

\listofchanges
\end{document}